\documentclass{article}

\usepackage[preprint]{neurips_2026}

\usepackage[utf8]{inputenc}
\usepackage[T1]{fontenc}

\usepackage{times}
\usepackage{pifont}
\usepackage{xcolor}
\usepackage{booktabs}
\usepackage{tabularx}
\usepackage{subcaption}
\usepackage{multirow}
\usepackage{xspace}
\usepackage{graphicx}
\usepackage{booktabs}
\usepackage{titlesec}
\titlespacing*{\paragraph}{0pt}{0pt}{0.5em}

\usepackage{booktabs}

\usepackage{subcaption}

\usepackage{graphicx}

\usepackage{makecell}

\usepackage{amsmath,amsfonts,bm}

\def\eqref#1{equation~\ref{#1}}

\def\1{\bm{1}}

\DeclareMathAlphabet{\mathsfit}{\encodingdefault}{\sfdefault}{m}{sl}
\SetMathAlphabet{\mathsfit}{bold}{\encodingdefault}{\sfdefault}{bx}{n}

\usepackage{hyperref}
\usepackage{url}

\newcommand{\name}{\textsc{PhylaFlow}\xspace}

\title{PhylaFlow: Hybrid Flow Matching in Billera–Holmes–Vogtmann Tree Space for Phylogenetic Inference}

\author{%
Yasha Ektefaie$^{1,*, \dagger}$ \quad
Leo Cui$^{2,*}$ \quad
Shrey Jain$^{3}$ \quad
Marinka Zitnik$^{2}$ \quad
Pardis Sabeti$^{4, \dagger}$ \\[0.75em]
$^{1}$Eric and Wendy Schmidt Center, Broad Institute of MIT and Harvard \\
$^{2}$Department of Biomedical Informatics, Harvard Medical School \\
$^{3}$Centennial High School \\
$^{4}$Infectious Disease and Microbiome Program, Broad Institute of MIT and Harvard \\[0.75em]
$^{*}$Equal contribution. \\
$^{\dagger}$Correspondence: \texttt{yektefai@broadinstitute.org}
}
\usepackage{geometry}

\begin{document}

\newgeometry{%
    textheight=9in,%
    textwidth=6.7in,%
    top=1in,%
    headheight=12pt,%
    headsep=25pt,%
    footskip=30pt%
  }%

\maketitle

\begin{abstract}

Phylogenetic trees are hybrid objects: branch lengths vary continuously, while topologies change discretely through edge contractions and expansions. Billera–Holmes–Vogtmann (BHV) tree space provides a canonical geometry for this structure, representing each resolved topology as a Euclidean orthant and topological changes as motion across shared lower-dimensional boundaries. This raises a natural question for generative modeling: can neural transport models learn meaningful motion through tree space? We introduce PhylaFlow, a hybrid flow-matching model for posterior-basin transport in BHV tree space. PhylaFlow is trained on BHV geodesic paths from random starting trees to short-run posterior samples, coupling continuous branch-length motion within orthants with learned boundary events and discrete topological transitions. We use Bayesian posterior-basin recovery as an operational test of this learned geometry: if the flow reaches meaningful regions of tree space, then finite-budget Bayesian refinement initialized from or guided by its terminal trees should recover posterior-supported topologies more efficiently. Across DS1–DS8 benchmarks, PhylaFlow substantially reduces initial Tree-KL relative to random, maximum-likelihood, maximum-parsimony, and UShER-based initializers. After finite-budget MrBayes refinement, direct PhylaFlow improves early and intermediate topology-recovery trajectories on most datasets, while a split-guided PhylaFlow-MCMC refinement obtains the strongest hard-case results. Compared with strong posterior-informed and neural controls, the best PhylaFlow variant outperforms short-warmup on seven of eight datasets and PhyloGFN on five of eight datasets under the same refinement budget. In a separate joint sequence-conditioned experiment, sequence embeddings steer posterior split recovery, although exact posterior topology recovery remains preliminary. These results show that hybrid flow matching can learn actionable transport in BHV tree space and provide a geometry-aware proposal mechanism for Bayesian phylogenetic inference.

\end{abstract}

\section{Introduction}
Phylogenetic inference is central to evolutionary biology, genomic surveillance, and the interpretation of biological sequence data\citep{kapli_phylogenetic_2020}. It allows us to reconstruct how pathogens spread, how variants diversify, and how species and genes have diverged across evolutionary time \citep{goig_ecology_2025, may_early_2024, hadfield_nextstrain_2018}. Yet despite its importance, phylogenetic inference is still often treated as a difficult combinatorial search problem over discrete tree topologies, coupled with continuous optimization or sampling over branch lengths \citep{kapli_phylogenetic_2020}.

Phylogenetic trees are not merely discrete combinatorial objects; they inhabit a hybrid geometric space that is neither purely continuous nor discrete. Branch lengths vary continuously, while topological changes occur through edge contractions and expansions\citep{bridgesbaysian}. Billera–Holmes–Vogtmann (BHV) tree space provides a canonical geometry for this hybrid object: each topology is an orthant, branch lengths are coordinates within that orthant, and topology changes occur by crossing shared lower-dimensional boundaries \citep{billera_geometry_2001}. By organizing tree topologies according to shared boundaries, BHV space turns tree search from a set of isolated combinatorial moves into a structured geometry in which nearby trees are meaningfully related. 

This geometric perspective suggests a natural question for generative modeling: can neural transport models learn meaningful motion through tree space? Bayesian phylogenetic inference offers a stringent way to test this question. Posterior mass over trees is often concentrated in multiple distant topological basins, and classical Markov chain Monte Carlo (MCMC) methods may spend substantial finite compute discovering these regions before likelihood-based refinement and calibration become effective\citep{alfaro_posterior_2006, lakner2008mcmc}. We therefore study whether learned BHV motion can transport arbitrary starting trees toward posterior-relevant basins.

\begin{figure}[t]
    \centering
    \includegraphics[width=0.8\linewidth]{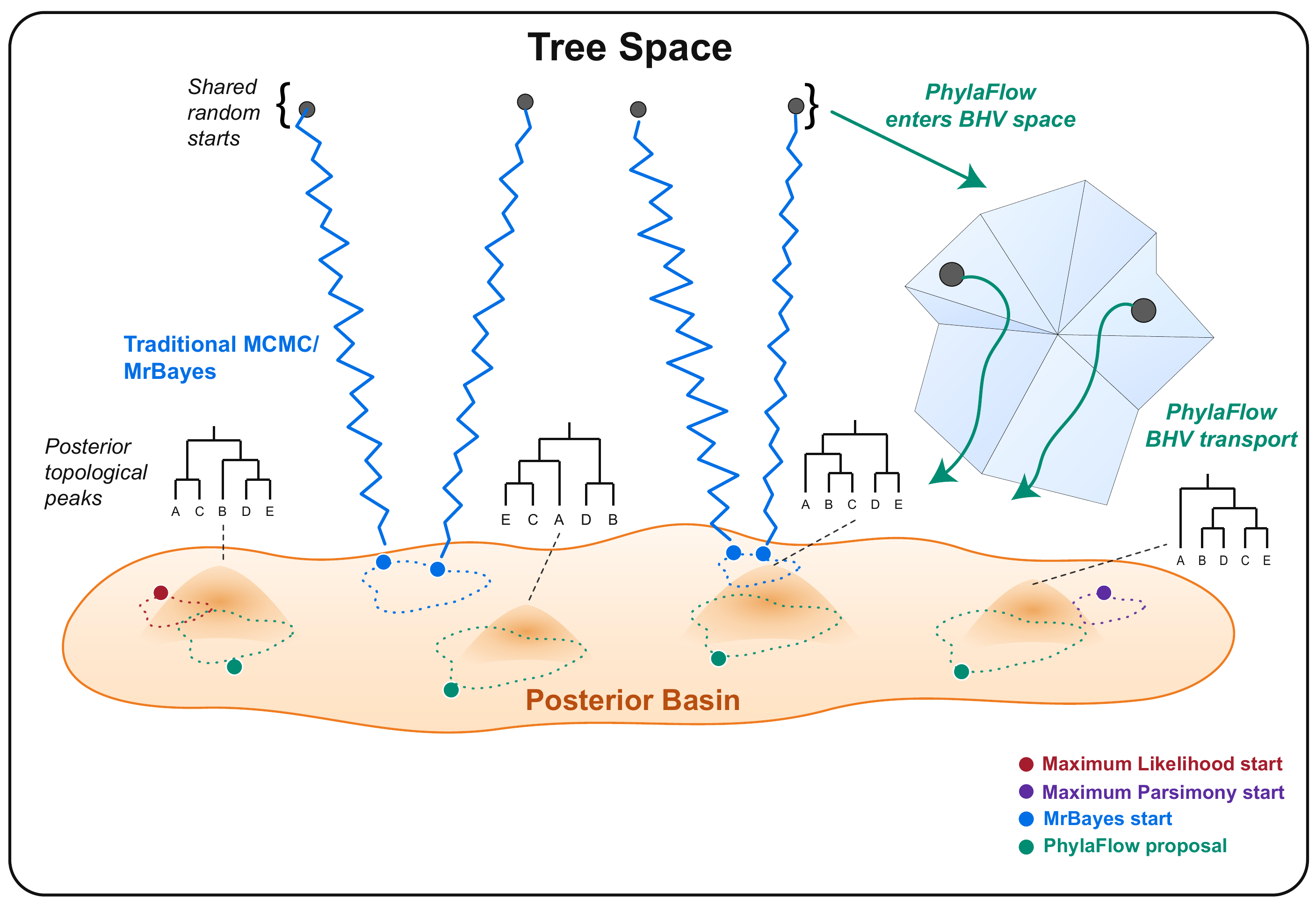}
    \caption{\textbf{Overview of PhylaFlow initialization in tree space.} Traditional MCMC/MrBayes chains initialized from shared random starts follow long exploratory trajectories before entering the posterior basin. PhylaFlow maps the same starts into BHV tree space and transports them to diverse candidate basin-entry proposals. In contrast, maximum likelihood and maximum parsimony provide single-point initializations. Distinct regions of the posterior basin correspond to different high-posterior tree topologies.}
    \label{fig:bhv-motion-overview}
\end{figure}

We propose \name, a hybrid flow model \citep{lipman2023flowmatching}for posterior-basin transport in BHV tree space, producing starting trees that enhance Bayesian phylogenetic inference. Rather than attempting to replace calibrated Bayesian inference, \name learns an amortized initializer: given a starting tree, it predicts tree-space motion toward regions supported by posterior samples. The model is trained from posterior-supervised trajectories between simple or heuristic starting trees and empirical posterior trees, learning continuous branch-length motion together with discrete topological transitions through BHV boundaries.

The \name formulation differs from existing neural phylogenetic models along both algorithmic and statistical axes. Some methods learn dataset-specific variational posterior distributions or sequential posterior samplers, often using alignment-based likelihood objectives during training \citep{zhang2018variational, Geophy, huelsenbeck_mrbayes_2001}. Others generate or optimize a single high-scoring tree from an alignment or sequence representation \citep{phyloformer, zhou2024phylogfn, duan_phylogen_2024}. In contrast, \name does not attempt to replace posterior sampling or point-tree optimization. It learns tree-to-tree motion in the intrinsic BHV geometry of phylogenetic trees: it transports random initial trees toward posterior-relevant basins. After training, this proposal step does not require a multiple-sequence alignment (MSA) or explicit likelihood evaluation; standard MCMC is then used for likelihood-based refinement and calibration.

Because a single proposed tree cannot establish whether the surrounding region of tree space has posterior support \citep{alfaro_posterior_2006, ronquist2012mrbayes}, we evaluate \name operationally through downstream Bayesian refinement. If \name terminal trees are closer to posterior-relevant regions, then standard MrBayes \cite{huelsenbeck_mrbayes_2001}, a widely used Bayesian phylogenetic MCMC sampler, initialized from those trees should reach lower topology error and higher likelihood earlier than chains initialized from conventional alternatives. We therefore measure Tree-KL, a divergence measure comparing sampled tree-topology distributions to a long-run reference posterior, and likelihood traces under matched MrBayes budgets. We compare \name starts against random trees, maximum likelihood starts \citep{IQTree}, maximum parsimony starts\citep{ronquist2012mrbayes, turakhia_ultrafast_2021}, and available neural tree-generation baselines \citep{zhou2024phylogfn}.

Across eight phylogenetic posterior benchmarks \citep{lakner2008mcmc} spanning diverse evolutionary regimes, \name produces better starting trees and improves downstream Bayesian refinement under matched compute budgets. The gains are most pronounced in the early and intermediate MCMC regime, supporting the view that learned BHV motion can reduce topology-space burn-in without replacing likelihood-based posterior calibration. We further find that learned split structure can guide post-hoc refinement in difficult cases, suggesting that the model captures actionable posterior geometry rather than merely producing isolated high-likelihood trees. Finally, we provide early evidence that this transport can be controlled by conditioning on sequence embeddings. Together, these results suggest that BHV geometry can be used not only to describe tree space, but to learn useful motion through it, providing an early example of differentiable modeling for phylogenetic inference

\section{Related Work}
\paragraph{Neural posterior approximation.}

A major line of work builds on variational Bayesian phylogenetic inference (VBPI), replacing MCMC with tractable variational distributions over tree topologies and branch lengths optimized against the phylogenetic likelihood and prior~\citep{zhang2018variational}. Subsplit Bayesian networks (SBNs)~\citep{zhang2018sbn}, ARTree~\citep{xie2023artree}, and ARTreeFormer~\citep{xie2025artreeformer} parameterize increasingly flexible topology distributions for posterior approximation or density estimation, while GeoPhy~\citep{Geophy} induces tree distributions from continuous coordinates and optimizes a likelihood-driven variational objective. These methods aim to represent a posterior distribution, often for a fixed dataset. \name instead learns posterior-informed tree-to-tree motion: its role is not to replace MCMC, but to initialize it closer to posterior-relevant regions.

\paragraph{Neural tree generation and likelihood-guided samplers.}

Other neural methods generate trees directly. Phyloformer predicts phylogenetic structure from multiple-sequence alignments, while PhyloGen uses genomic language-model features to generate and optimize tree topology and branch lengths without aligned sequence constraints or explicit evolutionary models~\citep{phyloformer,duan_phylogen_2024}. These methods provide natural neural start-tree baselines, but they primarily optimize or generate trees rather than learning posterior-basin transport. PhyloGFN is closer to posterior sampling: it learns a GFlowNet policy that sequentially constructs trees and, in its Bayesian version, is trained with a reward derived from the phylogenetic likelihood and branch-length prior~\citep{zhou2024phylogfn}. Thus PhyloGFN is a strong likelihood-guided neural sampler, whereas \name learns likelihood-free proposal-time motion in BHV space.

\section{Methods}
\subsection{Reframing Phylogenetics as an Optimal Transport Problem}

The goal of Bayesian phylogenetic inference is to characterize the posterior distribution over tree topologies and branch lengths given an observed sequence alignment,
\[
P(T,B \mid S,\theta),
\]
where \(T\) denotes the tree topology, \(B\) denotes branch lengths, \(S\) denotes the observed sequences, and \(\theta\) denotes additional model parameters or prior specifications, such as the substitution model, rate heterogeneity model, or tree prior \citep{alfaro_posterior_2006}. Many recent machine-learning approaches focus on a simplified setting in which inference is primarily conditioned on the sequence alignment \(S\), with such model parameters treated as fixed, implicit, or absorbed into the training distribution~\cite{xie2025phylovae, Geophy, phyloformer, ARTree}.

Estimating this posterior distribution is costly because it requires marginalizing over a combinatorially large space of possible tree topologies and continuous branch lengths\citep{lakner2008mcmc}:
\[
P(T,B \mid S,\theta) =
\frac{
P(S \mid T,B,\theta)\, P(T,B \mid \theta)
}{
\sum_{T' \in \mathcal{T}_n}
\int_{\mathcal{B}(T')}
P(S \mid T',B',\theta)\,
P(T',B' \mid \theta)\,
dB'
}.
\]
The denominator is the marginal likelihood, or model evidence. Computing it exactly is generally intractable because the number of possible tree topologies grows rapidly with the number of taxa, and each topology is associated with a continuous branch-length space. As a result, classical Bayesian phylogenetic methods approximate this posterior using Monte Carlo sampling, most commonly Markov chain Monte Carlo (MCMC), as in MrBayes and BEAST \citep{huelsenbeck_mrbayes_2001, bouckaert2014beast2}. Recent machine-learning approaches instead use amortized or variational inference methods to approximate the posterior more efficiently, for example by optimizing a tractable lower bound or surrogate objective \citep{zhang2018variational, zhang_improved_2020}; see Appendix~\ref{Appendix:Existing}.

We propose reframing phylogenetic inference as a \textit{transport problem} between a simple base distribution and the intractable posterior over trees and branch lengths. Specifically, \name learns a transport map that pushes forward samples from an initial distribution
\[
P_0(U,\zeta),
\]
where \(U\) represents a latent tree-structure variable and \(\zeta\) parameterizes positive branch lengths through \(B=\exp(\zeta)\), to an approximation of the target posterior distribution
\[
P_1(T,B \mid S,\theta).
\]

\subsection{BHV Tree Space as a Geometry for Local Tree Moves}

\begin{figure}[t]
    \centering
    \includegraphics[width=\linewidth]{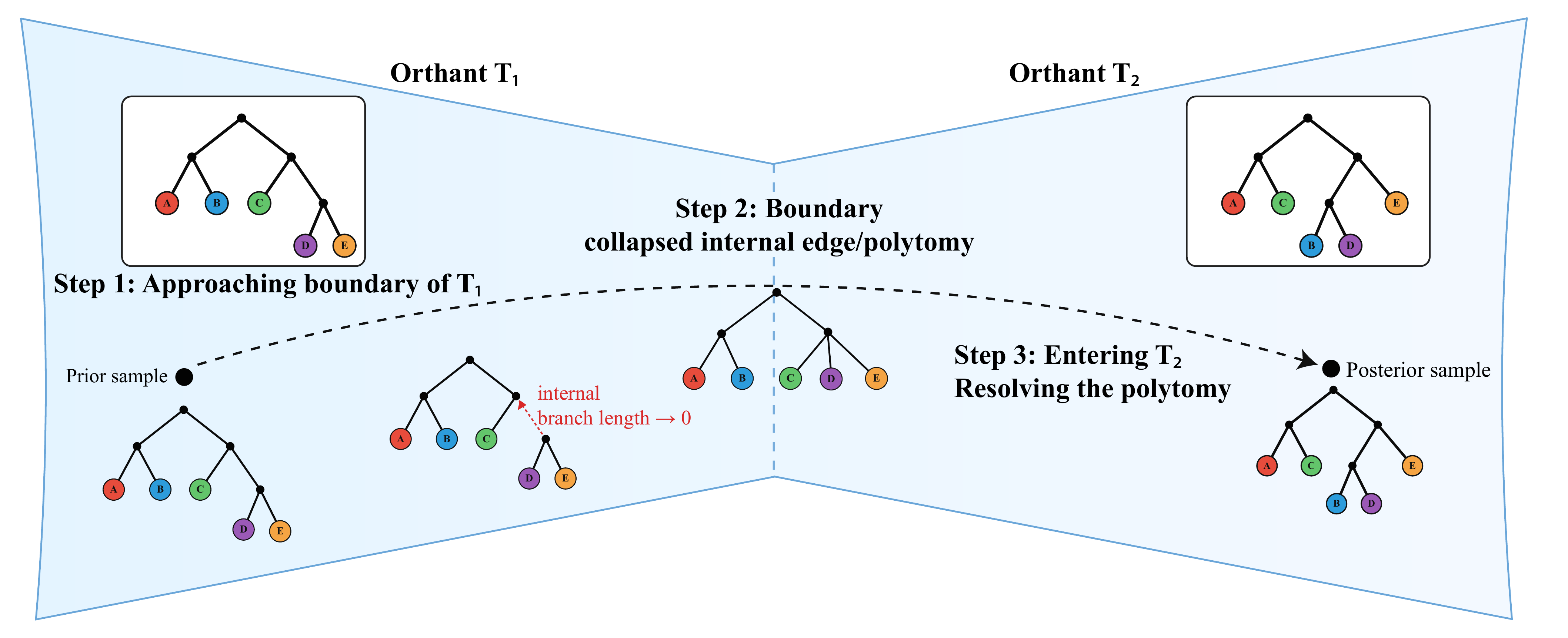}
    \caption{\textbf{Motion in BHV tree space.} Each orthant represents a fixed tree topology, with coordinates given by internal branch lengths. Moving within an orthant changes branch lengths without changing topology. When an internal branch contracts to zero, the path reaches a boundary corresponding to an unresolved tree; resolving that boundary in a different way moves the sample into a new topology. \name uses this geometry to combine continuous branch-length transport with learned discrete topology transitions.}
    \label{fig:bhv-motion}
\end{figure}

Tree topologies are combinatorial objects, so they do not admit ordinary Euclidean gradients. 
The Billera--Holmes--Vogtmann (BHV) space~\cite{BVH,bridgesbaysian} provides a useful geometric relaxation: for a fixed fully resolved unrooted topology \(T\) on \(N\) taxa, the internal branch lengths form a Euclidean orthant
\[
\mathcal{O}_T = \mathbb{R}_{\geq 0}^{N-3}.
\]
Different orthants are glued together along faces where one or more internal branch lengths are zero. 
Thus, continuous motion inside an orthant changes branch lengths while preserving topology, whereas crossing an orthant boundary corresponds to contracting edges, forming an unresolved tree, and resolving it into a different topology. 
We review the formal BHV metric and geodesic computation in Appendix~\ref{Appendix:BHV}.

In \name, we use BHV space as a local coordinate system for hybrid tree transport. Within the interior of a fixed topology \(T\), branch lengths evolve according to a time-dependent velocity field
\[
\dot{B}_t = V_\omega(T, B_t, t; \Phi(S)),
\qquad B_t \in \mathcal{O}_T^\circ ,
\]
which is trained by conditional flow matching on supervised branch-length velocities. This continuous component models how branch lengths should change while the topology remains fixed. Topology changes are handled separately as boundary events. When one or more internal branches approach zero, a boundary head predicts which branches vanish, and a discrete autoregressive transition model resolves the resulting polytomy into a neighboring topology. The resulting sampler is therefore hybrid: continuous flow transports samples within orthants, while learned discrete transitions move samples between orthants. 

\subsection{Model Architecture}

\name takes as input a tree state (topology and branch lengths), the current flow time, and sequence-derived embeddings to create node and edge tokens to input into a graph transformer \citep{kim_pure_2022}(Appendix~\ref{Appendix:Tokenization}). The hidden representation is then passed to three heads: the first-hit head, the velocity head, and the topology head (Figure~\ref{fig:model_arch}).

\textbf{Velocity head.}
The velocity head predicts a scalar branch-length velocity for each active edge:
\[
\hat{v}_{\omega,e}
=
g_{\mathrm{vel}}(h_e),
\qquad e \in E(T_t).
\]
This output parameterizes the continuous vector field within the current BHV orthant.

\textbf{First-hit head.}
The first-hit head predicts which active internal edges are expected to collapse first. 
\[
\hat{h}_{\omega,e}
=
g_{\mathrm{hit}}(h_e, k; c),
\qquad e \in E_{\mathrm{int}}(T_t).
\]
The resulting logits are used to identify the next boundary event, where one or more internal branches shrink to zero.

\textbf{Topology head.}
After a boundary event, the collapsed tree may contain an unresolved polytomy. The topology head resolves this polytomy by predicting a sequence of structured-subset actions:
\[
a_m
\sim
q_\omega
\left(
a_m
\mid
T_t, \ell_t, a_{<m}, t; \Phi(S), c
\right),
\]
where each \(a_m\) selects a starter pair, an additional member, a subset size, or a stopping decision. 

\paragraph{Two notions of time.}
\name distinguishes a continuous within-orthant time from a discrete phase index. Within an orthant, branch lengths evolve in a continuous local time \(\tau\in[0,1]\) used by the velocity field; integration along \(\tau\) advances \(\ell\) until the predicted boundary is reached. Across orthants, trajectories are organized into discrete phases indexed by \(k\in\{0,1,\dots,K\}\), where \(k\) increments each time a boundary event is taken. Representations for both and continuous times are integrated into node and edge representations, and continuous time is also passed into the first-hit head.
\paragraph{Start-tree case conditioning.}
Each training and inference path is initialized at a random starting tree (Appendix~\ref{Appendix:start_tree_generation}, ~\ref{Appendix:MetricEncoder}). We precompute a per-start embedding using a frozen metric encoder trained beforehand on tree-distance probes. This frozen embedding is injected into the first-hit head and the topology head receives it additively inside the structured-subset module. 

\subsection{Training Objective}

Each training example is derived from a supervised BHV geodesic between a sampled start tree and a target tree. Rather than training only on the endpoints, we decompose this geodesic into labeled tree states. A state is written \(x_t=(T_t,\ell_t)\), where \(T_t\) is the current topology and \(\ell_t\) are the active branch lengths. Each state is also associated with a within-orthant time \(\tau\), a discrete phase index \(k\), sequence-derived embeddings, and the frozen start-tree embedding \(c\). 

This decomposition produces two kinds of supervision. For states in the interior of an orthant, the topology is fixed and the model is trained to predict continuous branch-length motion within that orthant. Specifically, for each active edge \(e\in E(T_t)\), the velocity head is supervised with the geodesic velocity \(v^*_e\). The same state also provides supervision for the first-hit head: the target is the set \(H^*\subseteq E_{\mathrm{int}}(T_t)\) of active internal edges that collapse at the next boundary event. These intra-orthant examples therefore train the velocity field and the first-hit predictor. At a boundary, the geodesic specifies a transition between adjacent orthants. The end of the current orthant contains a set of death edges that collapse to zero, and the next orthant introduces a corresponding set of birth edges. We convert each boundary transition into a supervised autoregressive sequence of structured-subset actions. Starting from the unresolved polytomy obtained after the death edges collapse, each prefix \(a^*_{<m}\) defines a partially resolved tree, and the topology head is trained to predict the next action \(a^*_m\). These inter-orthant examples supervise the sequence of topology decisions that maps the boundary tree to the topology of the next orthant. The total loss combines these three supervised objectives:

\[
\mathcal{L}
=
\lambda_{\mathrm{vel}}\mathcal{L}_{\mathrm{vel}}
+
\lambda_{\mathrm{hit}}\mathcal{L}_{\mathrm{hit}}
+
\lambda_{\mathrm{top}}\mathcal{L}_{\mathrm{top}}
+
\lambda_{\mathrm{anchor}}\mathcal{L}_{\mathrm{anchor}}.
\]

Here, \(\mathcal{L}_{\mathrm{vel}}\) matches predicted and supervised branch-length velocities on intra-orthant states, \(\mathcal{L}_{\mathrm{hit}}\) trains the first-hit head to identify the next collapsing internal edge set, and \(\mathcal{L}_{\mathrm{top}}\) is the autoregressive loss for the inter-orthant states. We additionally include an auxiliary anchor loss on states from fixed supervised start--target paths to stabilize the direction of edge contraction. Once trained these heads are used in a deterministic sampling algorithm to produce movement in BHV space (Appendix~\ref{Appendix:Sampling}). Detailed definitions and implementation choices are provided in Appendix~\ref{Appendix:TrainingObjective}.

\section{Results}
We ask whether learned transport in BHV tree space can move unseen starting trees toward posterior-relevant regions of tree space. We evaluate this with complementary diagnostics that capture both topological agreement and likelihood plausibility. TreeKL measures agreement between the sampled topology distribution and an independent long-run \textsc{MrBayes} posterior, rewarding methods that recover posterior-supported topologies and multiple topological modes rather than only a single high-likelihood tree. Split-KL provides a coarser marginal view by comparing posterior split frequencies. Finally, an optimized-likelihood gap checks whether sampled topologies lie in a plausible high-likelihood region under the alignment. Formal definitions are given in Appendix~\ref{Appendix: Metrics}.

We organize the results around four questions: 
(i) does learned BHV transport move unseen starts toward posterior-supported topologies; 
(ii) does this remain true under stronger posterior-informed and neural controls; 
(iii) can the transport be conditioned by sequence embeddings; and 
(iv) what geometric motion does the model learn? Component ablations are reported in Appendix~\ref{Appendix:ablations}.

\subsection{Can \name learn meaningful BHV motion?}

\paragraph{Experimental Setup}For each benchmark dataset DS1--DS8 \citep{lakner2008mcmc}, we use two independent \textsc{MrBayes} runs: a short run used only to construct training targets for \name, and a longer run used as the reference posterior for evaluation \citep{ARTree}. Let $\widehat{\pi}^{\mathrm{short}}_d$ denote the empirical topology distribution from the short run for dataset $d$. We train \name on BHV geodesic paths from random starting trees to target posterior topologies sampled according to $\widehat{\pi}^{\mathrm{short}}_d$, so topologies appearing more frequently in the short posterior run are sampled more often as transport targets.

At evaluation time, \name is applied to unseen random starts (Appendix~\ref{Appendix:start_tree_generation}), and the resulting trees are used to initialize independent \textsc{MrBayes} chains. We compare against classic baselines in Phylogenetics (Appendix~\ref{Appendix:classic_methods}) under the same downstream \textsc{MrBayes} configuration and the same $10^5$-generation budget. We exclude sequence-derived embeddings from these experiments. At inference time, \name receives no sequence or likelihood information; it is evaluated only on whether learned BHV transport generalizes to unseen start trees.

\paragraph{Topology-distribution agreement.}

Table~\ref{tab:tree_kl_initial_final} reports initial TreeKL before MCMC and final TreeKL after the $10^5$-generation \textsc{MrBayes} run. Before MCMC, \name obtains the lowest TreeKL on all eight datasets, whereas most standard initializers have little or no overlap with the reference posterior topology support. This shows that learned BHV transport moves unseen random starts substantially closer to posterior-supported topological regions before any likelihood-based MCMC correction.

After standard finite-budget \textsc{MrBayes} refinement, direct \name initialization remains the best non-guided initialization on DS1, DS2, DS4, DS5, and DS8. On DS3, DS6, and DS7, standard \textsc{MrBayes} refinement from maximum-parsimony or related starts gives lower final TreeKL. However, the checkpoint trajectories show that \name can reach stronger topology-distribution agreement earlier in the chain than is preserved at the final checkpoint. On DS3, \name reaches TreeKL $2.134$ by 20K generations, but ends at $2.539$ by 100K. On DS6, \name reaches TreeKL $4.814$ by 40K generations, but ends at $5.042$ by 100K. These non-monotone trajectories suggest that standard local \textsc{MrBayes} proposals do not always preserve or exploit the basin structure supplied by the learned initialization under a finite budget.

We therefore also evaluate \name as a proposal guide in a PhylaFlow-MCMC refinement scheme. PhylaFlow-MCMC constructs a guide bank of \name-generated trees, estimates split weights from this bank, and uses these weights to bias local NNI proposals toward PhylaFlow-supported splits (Appendix~\ref{Appendix:phylaflowmcmc}). PhylaFlow-MCMC obtains the best final TreeKL on DS3, DS5, DS6, DS7, and DS8, while direct \name initialization remains best on DS1, DS2, and DS4. Longer standard \textsc{MrBayes} runs did not eliminate this gap in the tested hard cases: on DS3, the best baseline after one million steps gives TreeKL $2.365$, compared to $0.117$ for PhylaFlow-MCMC; on DS6, baseline one million steps gives TreeKL $5.006$, compared to $4.479$ for PhylaFlow-MCMC (Appendix~\ref{app:mrbayes-long-chain}). These results suggest that \name provides posterior-relevant proposal information that standard \textsc{MrBayes} may not discover efficiently from many starts.

\begin{table}[ht]
\centering
\small
\setlength{\tabcolsep}{4.2pt}
\renewcommand{\arraystretch}{1.08}
\caption{Initial and final Tree-KL (lower is better) relative to the long-run posterior topology distribution. Panel A reports the Tree-KL of the supplied initial topology distribution before MCMC. Panel B reports the final Tree-KL after finite-budget MCMC run. }
\label{tab:tree_kl_initial_final}
\begin{tabular}{lrrrrrrrr}
\toprule
Method & DS1 & DS2 & DS3 & DS4 & DS5 & DS6 & DS7 & DS8 \\
\midrule
\multicolumn{9}{l}{\textbf{Initial Tree-KL, before MCMC}} \\
\midrule
Random       & 15.442 & 16.169 & 17.609 & 17.042 & 15.406 & 16.852 & 16.417 & 16.407 \\
MP           & 15.442 & 16.169 & 12.637 & 17.042 & 15.406 & 16.852 & 14.357 & 16.407 \\
IQ-TREE      & 11.775 & 16.169 & 17.609 & 17.042 & 15.406 & 16.852 & 16.417 & 16.407 \\
UShER-s      & 15.442 & 16.169 & 17.609 & 17.042 & 15.406 & 16.852 & 16.417 & 16.407 \\
UShER-ms     & 15.442 & 16.169 & 17.609 & 17.042 & 15.406 & 16.852 & 16.417 & 16.407 \\
\midrule
PhylaFlow    & \textbf{2.728} & \textbf{0.505} & \textbf{2.941} & \textbf{7.631} & \textbf{6.020} & \textbf{8.005} & \textbf{11.345} & \textbf{9.557} \\
\midrule
\multicolumn{9}{l}{\textbf{Final Tree-KL, after finite-budget MCMC}} \\
\midrule
Random       & 0.403 & 0.072 & 2.422 & 0.263 & 4.274 & 5.108 & 2.628 & 0.905 \\
MP           & 0.430 & 0.071 & 2.106 & 0.277 & 4.215 & 4.972 & 2.297 & 0.987 \\
IQ-TREE      & 4.051 & 0.067 & 2.406 & 0.184 & 6.790 & 5.479 & 2.651 & 0.989 \\
UShER-s      & 0.586 & 0.086 & 2.516 & 0.294 & 4.823 & 5.087 & 2.678 & 0.939 \\
UShER-ms     & 0.435 & 0.082 & 2.650 & 0.273 & 4.136 & 5.057 & 2.456 & 0.821 \\
\midrule
PhylaFlow    & \textbf{0.150} & \textbf{0.049} & 2.539 & \textbf{0.169} & \textbf{3.448} & \textbf{5.042} & 2.468 & \textbf{0.787} \\
PhylaFlow-MCMC & 0.304 & 0.139 & \textbf{0.117} & 0.514 & \textbf{1.808} & \textbf{4.479} & \textbf{1.071} & \textbf{0.612} \\
\bottomrule
\end{tabular}
\end{table}

\paragraph{Likelihood-based posterior-basin diagnostic.}

TreeKL measures posterior topology-distribution agreement, but it does not by itself verify that sampled topologies lie in a plausible likelihood region. To investigate this we compare the median log-likelihood of generated topologies to that of the posterior (Appendix \ref{Appendix:likelihood_diagnostic}). Table~\ref{tab:init_tail_likelihood_gap} shows that before MCMC, random starts are far outside the posterior-core likelihood regime, with gaps of thousands to tens of thousands of log-likelihood units. In contrast, \name proposes initial topologies with small optimized-likelihood gaps on most datasets and obtains the smallest initializer-only gap on DS2, DS3, DS4, DS5, and DS8. IQ-TREE remains strongest on DS1, DS6, and DS7, as expected for a likelihood-optimized initializer. After $10^5$ generations of \textsc{MrBayes}, the likelihood gaps become small for all methods. This indicates that \name improves posterior-topology occupancy without sending chains into likelihood-poor regions.

\subsection{Does learned BHV motion remain useful under stronger controls?}

The preceding experiments test the geometric question: can learned BHV transport move unseen random trees toward posterior-relevant basins? We next compare against two stronger controls with more direct access to posterior or likelihood information. The short-warmup baseline initializes directly from the same kind of short \textsc{MrBayes} posterior sketch used to supervise \name. PhyloGFN~\citep{zhou2024phylogfn} is a state-of-the-art likelihood-trained neural sampler.

Table~\ref{tab:strong_baseline_stress} shows that these controls are strong, but do not eliminate the value of learned BHV motion. PhyloGFN gives the best 100K Tree-KL on DS1, DS4, and DS8, while direct \name is best on DS2. On the harder finite-budget cases, \name-MCMC is best on DS3, DS5, DS6, and DS7. To further isolate the role of initialization from the downstream MCMC procedure, we also evaluate a PhyloGFN-MCMC variant that uses the same refinement procedure as \name-MCMC, but initializes from PhyloGFN samples. This variant improves over \name-MCMC on DS3, but \name-MCMC remains better on DS5, DS6, and DS7, the harder posterior distributions in this group (Appendix~\ref{Appendix:PhyloGFNMCMC}).

Cost also matters. Table~\ref{tab:setup_runtime_strong_controls} reports method-specific setup time before the shared $10^5$-generation \textsc{MrBayes} refinement. Short-warmup is cheapest, but often less accurate. PhyloGFN is strong, but its estimated setup time grows sharply on the larger benchmarks, reaching 43--48 hours on DS7--DS8. In contrast, even when \name is charged for the short-run supervision used to construct training targets, its setup time stays between 10--26 hours on DS2--DS8 and totals 140.23 hours across all datasets, compared with 235.92 hours for PhyloGFN (Appendix~\ref{Appendix:run_time_estimate}). This test therefore supports a quality--cost conclusion: learned BHV transport gives a comparatively efficient path to proposal-useful posterior geometry.

\begin{table}[t]
\centering
\small
\setlength{\tabcolsep}{4.2pt}
\renewcommand{\arraystretch}{1.06}
\caption{ Tree-KL after $10^5$ generations of finite-budget refinement. 
Best classical is the best result from the baselines in Table~\ref{tab:tree_kl_initial_final}. Short-warmup initializes directly from the short MrBayes posterior sketch used to supervise \name.}
\label{tab:strong_baseline_stress}
\begin{tabular}{lrrrrrrrr}
\toprule
Method & DS1 & DS2 & DS3 & DS4 & DS5 & DS6 & DS7 & DS8 \\
\midrule
Best classical 
& 0.403 & 0.067 & 2.106 & 0.184 & 4.136 & 4.972 & 2.297 & 0.821 \\
short-warmup 
& 0.164 & 0.061 & 1.956 & 0.206 & 2.948 & 4.949 & 1.990 & 0.502 \\
PhyloGFN 
& \textbf{0.103} & 0.051 & 1.983 & \textbf{0.151} & 2.950 & 5.337 & 2.353 & \textbf{0.313} \\
\midrule
\name 
& 0.150 & \textbf{0.049} & 2.539 & 0.169 & 3.448 & 5.042 & 2.468 & 0.787 \\
\name-MCMC 
& 0.304 & 0.139 & \textbf{0.117} & 0.514 & \textbf{1.808} & \textbf{4.479} & \textbf{1.071} & 0.612 \\
\bottomrule
\end{tabular}
\end{table}

\subsection{Can \name be conditioned on sequence embeddings?}

To test whether \name can learn a sequence-conditioned transport map, we train a single checkpoint jointly on DS1--DS8 and condition generation on dataset-specific embeddings of the input alignment. We use Phyla embeddings~\cite{ektefaie2026evolutionary} (Appendix~\ref{Appendix:condit:phyla}). This setting is substantially harder than the per-dataset regime used above: one shared checkpoint must represent multiple posterior landscapes with different taxon sets and alignments. 

The joint conditioned checkpoint achieves low split-KL under native conditioning, indicating that it recovers much of the dataset-specific posterior split mass. However, its exact-topology Tree-KL before refinement is substantially worse than dataset-specific \name, suggesting that the model often identifies posterior-relevant splits without yet assigning accurate joint mass to complete topologies (Table~\ref{Appendix:condit:kl}). The conditioned model gives useful posterior-region starts, but does not yet achieve state-of-the-art amortized inference across datasets.

To test whether generation is genuinely controlled by the sequence conditioning signal, we perform a cross-conditioning stress test. For each target dataset, we fix the checkpoint, starting topologies, posterior reference, and scoring procedure, and vary only the Phyla embedding bank supplied to the sampler. The diagonal condition uses the target dataset's own embedding bank, while off-diagonal conditions swap in another dataset's embedding bank. Native conditioning achieves the lowest split-KL for every target with at least one valid off-diagonal swap. Averaged over seven targets, native conditioning gives split-KL $0.236$, whereas the best valid off-diagonal swap gives split-KL $3.596$, a mean native advantage of $3.359$ (Table~\ref{Appendix:condit:crosstable}). Because only the conditioning bank changes within each target row, this diagonal dominance shows that the joint checkpoint is not merely learning an unconditional prior over plausible trees. Its flow is actively routed by the sequence-derived conditioning signal.

\subsection{What motion does \name learn?}
\label{Appendix:PhylaFlowMotion}

\begin{figure*}[t]
\centering
\includegraphics[width=\textwidth]{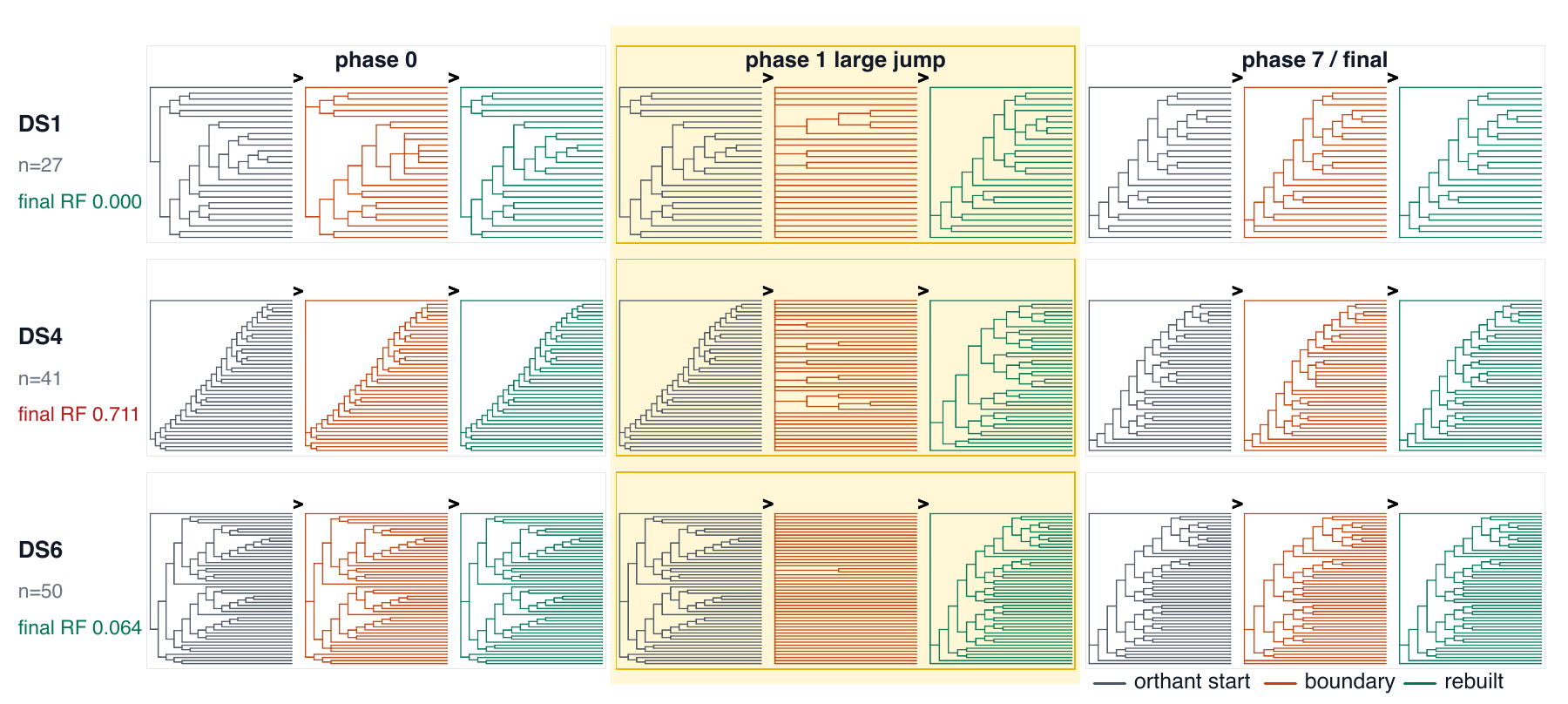}
\caption{
Partial representative topology trajectories learned by \name. Each phase shows a legal BHV transition from orthant start to boundary to rebuilt tree. The highlighted phase~1 shows the typical large collapse--rebuild step, followed by smaller refinements. RF denotes Robinson--Foulds distance to the reference tree.}
\label{fig:topology-trajectories}
\end{figure*}

To examine this behavior, we inspected 50 test-time trajectories. Figure~\ref{fig:topology-trajectories} shows three representative examples from this inspection. These trajectories suggest that \name does not simply reproduce the BHV geodesic between the start and target trees. Instead, it learns a distinct amortized motion through tree space.

The learned motion is typically front-loaded. In the first phase, \name often collapses many internal splits simultaneously, producing a low-dimensional boundary tree, and then rebuilds the tree by inserting many new short splits into a different orthant. Subsequent phases are smaller: they usually add, remove, or rearrange only a few splits, acting as local refinements rather than large topological jumps. This pattern is visible in Figure~\ref{fig:topology-trajectories}: the highlighted phase-1 transition performs the dominant destruction and reconstruction step, while the final phases refine or stabilize the resulting topology.

\section{Discussion}
This work proposes a hybrid approach to phylogenetic inference in which neural transport improves posterior-basin discovery while likelihood-based MCMC is retained for statistical refinement. \name learns geometry-aware motion in BHV tree space, where branch-length changes and topology transitions are represented within a shared structure. Across DS1–DS8, \name moved random starting trees toward posterior-relevant regions before MCMC refinement, improving agreement with posterior split structure across all benchmarks.

A key challenge in BHV space is that useful movement between topologies requires coordinated behavior at orthant boundaries, where internal edges contract and new topological structures emerge. \name was able to learn transport dynamics that produced useful topology and branch-length changes from unseen starting trees. These results suggest that actionable motion in BHV space is learnable.

\name is therefore best understood as an initializer or proposal mechanism rather than a replacement for Bayesian inference. Its proposals maintained plausible likelihood profiles while improving posterior-topology agreement, suggesting that the model captured reusable posterior structure rather than simply producing isolated high-scoring trees. The split-informed refinement results further support this interpretation. The sequence-conditioned experiments provide an initial step toward amortized phylogenetic inference, showing that sequence-derived embeddings can steer transport toward dataset-relevant split structure.

Several limitations remain. First, the current sampling procedure can be slow because autoregressive topology updates require repeated model calls, especially when many rearrangement or merge steps are needed. More efficient batching, parallel proposal generation, or non-autoregressive approximations could substantially improve scalability. Second, sequence-conditioned performance remains weaker than dataset-specific training, indicating that cross-dataset generalization is still limited. Third, the current implementation was not optimized for training efficiency, particularly in how labels and intermediate tree states were loaded. Finally, while frozen sequence embeddings provided useful conditioning signal, end-to-end training of the sequence and tree components may learn representations better aligned with phylogenetic transport.

Overall, \name shows that differentiable, geometry-aware models can learn useful motion in phylogenetic tree space while complementing existing likelihood-based inference methods. More broadly, these results suggest a path toward amortized phylogenetic models that learn reusable patterns of evolutionary structure across datasets.

\section*{Code Availability}

Data can be found here \url{https://zenodo.org/records/20297912} and code can be found here \url{https://github.com/yashaektefaie/PhylaFlow}.

\begin{ack}
Y.E. is supported by the Eric and Wendy Schmidt Center at Broad Institute.

The authors would like to acknowledge Patrick Varilly for his critical feedback while drafting this manuscript.
\end{ack}

\bibliographystyle{plainnat}
\bibliography{references, leo}

\appendix

\section{Appendix}
\subsection{Existing approaches to phylogenetics}
\label{Appendix:Existing}
Bayesian phylogenetic inference aims to characterize the posterior distribution
over tree topologies and branch lengths,
\[
p(T,B \mid S,\theta)
=
\frac{
p(S \mid T,B,\theta)\,p(T,B \mid \theta)
}{
p(S \mid \theta)
},
\]
where \(T\) denotes the tree topology, \(B\) denotes the branch lengths, \(S\)
denotes the observed sequence alignment, and \(\theta\) denotes additional
evolutionary-model parameters. The marginal likelihood,
\[
p(S \mid \theta)
=
\sum_{T' \in \mathcal{T}_n}
\int
p(S \mid T',B',\theta)\,p(T',B' \mid \theta)\,dB',
\]
is generally intractable because the number of possible tree topologies grows
combinatorially with the number of taxa and each topology has a continuous
branch-length space.

Classical Bayesian phylogenetic methods therefore approximate this posterior
using Monte Carlo sampling, most commonly Markov chain Monte Carlo (MCMC)
\citep{huelsenbeck_mrbayes_2001}. Given posterior samples
\[
(T_i,B_i) \sim p(T,B \mid S,\theta), \qquad i=1,\ldots,N,
\]
expectations under the posterior can be approximated as
\[
\mathbb{E}_{p(T,B \mid S,\theta)}[f(T,B)]
\approx
\frac{1}{N}\sum_{i=1}^{N} f(T_i,B_i).
\]
For example, posterior split frequencies, topology probabilities, or summary
statistics of branch lengths can be estimated by choosing an appropriate
function \(f(T,B)\). In this framework, the likelihood
\(p(S \mid T,B,\theta)\) is used to evaluate candidate trees during sampling,
while the final output is an empirical approximation to the posterior
distribution.

Variational phylogenetic methods take a different approach. Rather than drawing
samples from the posterior using a Markov chain, they introduce a parameterized
distribution \(q_\phi(T,B \mid S)\) and optimize it to approximate the true
posterior \citep{zhang2018variational}. A common objective is to minimize the
Kullback--Leibler divergence
\[
\phi^*
=
\arg\min_\phi
D_{\mathrm{KL}}
\left(
q_\phi(T,B \mid S)
\,\|\, 
p(T,B \mid S,\theta)
\right).
\]
Equivalently, variational inference maximizes the evidence lower bound (ELBO),
\[
\mathcal{L}_{\mathrm{ELBO}}(\phi)
=
\mathbb{E}_{q_\phi(T,B \mid S)}
\left[
\log p(S \mid T,B,\theta)
\right]
-
D_{\mathrm{KL}}
\left(
q_\phi(T,B \mid S)
\,\|\,
p(T,B \mid \theta)
\right).
\]
The first term rewards trees that explain the observed alignment well under the
phylogenetic likelihood, while the second term regularizes the variational
distribution toward the prior over topologies and branch lengths. Recent neural
and variational approaches use increasingly expressive parameterizations of
\(q_\phi\), including structured topology distributions and generative models
over trees \citep{zhang2018variational,duan_phylogen_2024, phyloformer, zhou2024phylogfn, xie2025phylovae}

\section{The Billera--Holmes--Vogtmann (BHV) Metric}
\label{Appendix:BHV}

The Billera--Holmes--Vogtmann (BHV) space~\cite{BVH} provides a geometric representation of phylogenetic trees with a fixed leaf set. 
For \(N\) taxa, each fully resolved unrooted topology \(T\) corresponds to a closed orthant
\[
\mathcal{O}_T = \mathbb{R}_{\ge 0}^{N-3},
\]
whose coordinates are the internal branch lengths of \(T\). 
Points in the interior of \(\mathcal{O}_T\) represent trees with topology \(T\) and strictly positive internal branch lengths. 
Boundary faces, where one or more internal branch lengths are zero, correspond to unresolved trees and are identified with the corresponding faces of other orthants. 
The resulting space
\[
\mathcal{T}_N = \mathrm{BHV}_N
\]
is a piecewise-Euclidean CAT(0) metric space. 
Consequently, any two trees \(x_1,x_2\in\mathcal{T}_N\) are connected by a unique geodesic \(\Gamma_{x_1,x_2}\), and the BHV distance \(d_{\mathrm{BHV}}(x_1,x_2)\) is the length of this geodesic.

Within a single orthant, \(d_{\mathrm{BHV}}\) reduces to the Euclidean distance between branch-length vectors. 
When two trees have different topologies, the geodesic is a piecewise-linear path through a sequence of orthants. 
Along this path, edges present in the first tree but incompatible with the target topology are contracted to zero, the path crosses lower-dimensional faces corresponding to unresolved trees, and edges of the target topology are expanded. 
Edges shared by the two endpoint trees remain present along the geodesic, with their lengths varying continuously.

\subsection{Efficient Computation of BHV Geodesics}

Although the number of tree topologies grows super-exponentially with \(N\), BHV geodesics can be computed in polynomial time. 
Owen and Provan~\cite{owen2011fast} gave an \(O(N^4)\)-time algorithm for computing the BHV geodesic between two trees. 
The algorithm represents each internal edge by its split of the taxon set and characterizes the geodesic by an ordered support sequence
\[
(A_1,B_1),\ldots,(A_k,B_k),
\]
where \(A_i\) is a set of edges contracted from the first tree and \(B_i\) is a set of edges expanded toward the second tree.

Let \(x_1=(T_1,\ell_1)\) and \(x_2=(T_2,\ell_2)\), with internal split sets \(E_1\) and \(E_2\). 
Let
\[
E_{\mathrm{common}} = E_1 \cap E_2
\]
denote the splits shared by the two trees. 
For a set of edges \(A\subseteq E_1\), define
\[
\|\ell_1(A)\|_2
=
\left(\sum_{e\in A}\ell_1(e)^2\right)^{1/2},
\]
and define \(\|\ell_2(B)\|_2\) analogously for \(B\subseteq E_2\). 
Given the geodesic support sequence \(\{(A_i,B_i)\}_{i=1}^k\), the BHV distance is
\[
d_{\mathrm{BHV}}(x_1,x_2)^2
=
\sum_{e\in E_{\mathrm{common}}}
\big(\ell_1(e)-\ell_2(e)\big)^2
+
\sum_{i=1}^k
\left(
\|\ell_1(A_i)\|_2
+
\|\ell_2(B_i)\|_2
\right)^2 .
\]
The first term accounts for length changes on shared internal edges. 
The second term accounts for the contraction and expansion of incompatible edge sets along the geodesic. 
Owen and Provan compute the support sequence by solving a sequence of polynomial-time combinatorial subproblems on incompatibility graphs. From the geodesic we derive all labels for training \name. We precalculate all geodesics used in training to avoid the computational overhead.

\section{Extended Methods}

\begin{figure}[t]
    \centering
    \includegraphics[width=\linewidth]{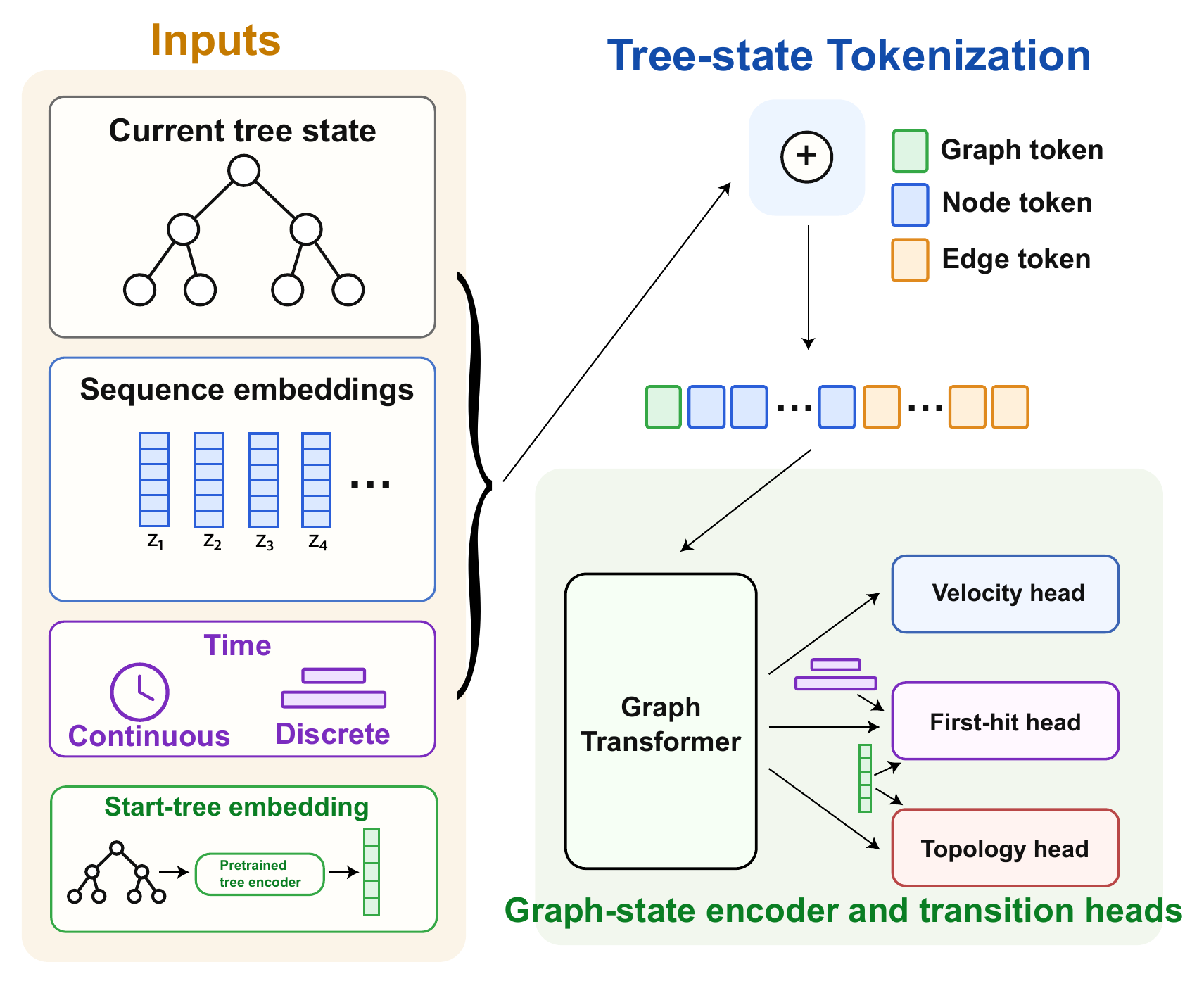}
    \caption{\textbf{PhylaFlow graph-state encoder and transition heads.} PhylaFlow represents the current tree, sequence information, time, and start-tree conditioning as graph, node, and edge tokens. A shared graph transformer encodes this tokenized tree state, after which separate heads predict continuous branch-length velocities, boundary first-hit events, and autoregressive topology transitions. This parameterization couples continuous motion within a tree-space orthant with discrete transitions across orthant boundaries.
    }
    \label{fig:model_arch}
\end{figure}

\subsection{\name Input and Tokenization}
\label{Appendix:Tokenization}
Each tree state \(x_t = (T_t, \ell_t)\) is represented as a sequence of node tokens and edge tokens. For a tree with node set \(V(T_t)\) and edge set \(E(T_t)\), the token sequence has length
\[
M = |V(T_t)| + |E(T_t)|.
\]
Node and edge tokens have different structural content and are constructed separately. For a node \(v \in V(T_t)\),
\[
x_v
=
e^{\mathrm{node}}(v)
+
e^{\mathrm{seq}}(v)
+
e^{\mathrm{type}}_{\mathrm{node}}
+
e^{\mathrm{lap}}(v,v),
\]
and for an edge \(e=(u,v) \in E(T_t)\),
\[
x_e
=
e^{\mathrm{edge}}(e)
+
e^{\mathrm{branch}}(\ell_e)
+
e^{\mathrm{seq}}(e)
+
e^{\mathrm{type}}_{\mathrm{edge}}
+
e^{\mathrm{lap}}(u,v),
\]
where \(e^{\mathrm{node}}\) and \(e^{\mathrm{edge}}\) embed the node and edge categorical types, \(e^{\mathrm{branch}}\) encodes the branch length, \(e^{\mathrm{type}}_{\{\mathrm{node},\mathrm{edge}\}}\) is a learned token-type embedding distinguishing node tokens from edge tokens, \(e^{\mathrm{seq}}\) is the sequence-derived conditioning described below, and \(e^{\mathrm{lap}}\) is a Laplacian positional encoding of the tree graph: a learned linear projection of \([\mathrm{lap}(u)\,\|\,\mathrm{lap}(v)]\), where \(\mathrm{lap}(\cdot)\) collects the leading non-trivial eigenvectors of the tree's graph Laplacian. A separate sinusoidal index positional embedding is also added to each token, computed independently over the node and edge sequences. 

\paragraph{Sequence conditioning} \name conditions tree generation on sequence-derived embeddings,
\[
\Phi(S) = \{z_1, \dots, z_N\},
\]
where \(z_i\) represents the embedding of taxon \(i\).  

For leaf tokens, the corresponding sequence embedding is projected into the model dimension and added directly to the leaf token:
\[
x_i^{\mathrm{leaf}}
\leftarrow
x_i^{\mathrm{leaf}} + W_{\mathrm{leaf}} z_i.
\]

For edge tokens, each internal edge defines a bipartition of the taxa. Let \(A_e\) and \(A_e^c\) denote the two sides of the split induced by edge \(e\). We compute a split-level conditioning vector by averaging the sequence embeddings on each side of the bipartition and passing their concatenation through an MLP:
\[
z_e^{\mathrm{split}}
=
\mathrm{MLP}_{\mathrm{split}}
\left(
\left[
\frac{1}{|A_e|}\sum_{i \in A_e} z_i
\ ;
\ 
\frac{1}{|A_e^c|}\sum_{i \in A_e^c} z_i
\right]
\right).
\]

This split embedding is added to the corresponding edge token,
\[
x_e^{\mathrm{edge}}
\leftarrow
x_e^{\mathrm{edge}} + z_e^{\mathrm{split}}.
\]

After tokenization and conditioning, a learnable graph-level token is prepended to the token sequence. The final input to the encoder is
\[
X_t =
[x_{\mathrm{graph}}, x_1, \dots, x_M].
\]
This sequence is passed through a stack of prenorm transformer encoder layers with multi-head self-attention. The encoder produces contextualized representations for the graph token, node tokens, and edge tokens that are routed to task-specific prediction heads.
\subsection{Training Objective}
\label{Appendix:TrainingObjective}

Training is performed from supervised tree-transition paths.  Each training
state is a tree
\[
x_t=(T_t,B_t),
\]
where $T_t$ is the current topology and $B_t$ are its active branch lengths. We train the model to predict the local continuous motion inside an orthant and the discrete events that determine when the trajectory reaches a boundary. For each sampled state along a path, the supervision consists of a target
branch-length velocity $v_t^\star$, a first-hit collapse set $C^\star$, boundary timing targets, and a sequence of autoregressive topology actions $a_m^\star$ that resolves the polytomy at each boundary into the next topology along the path.

\paragraph{Continuous velocity loss.}
Within a fixed topology $T_t$, the active BHV orthant is Euclidean.  The velocity
head predicts branch-length derivatives on the active edges:
\[
\hat v_\omega(x_t,t)
=
\left(\hat v_{\omega,e}(x_t,t)\right)_{e\in E(T_t)} .
\]
We regress these predictions against path-derived target velocities:
\[
\mathcal L_{\mathrm{vel}}
=
\mathbb E
\left[
\sum_{e\in E(T_t)}
w_e
\left(
\hat v_{\omega,e}(x_t,t)
-
v_{t,e}^{\star}
\right)^2
\right].
\]
Thus, the continuous component is a local flow-matching objective inside the
currently active orthant.

\paragraph{First-hit collapse loss.}
Topology changes occur when one or more active internal edges shrink to zero.
For each contracting edge, the target path induces a collapse time
\[
\tau_e^\star
=
\frac{B_{t,e}}{[-v_{t,e}^\star]_+},
\qquad
[-v]_+ = \max(-v,0).
\]
The target first-hit set is
\[
C^\star
=
\left\{
e :
\tau_e^\star
\le
\min_j \tau_j^\star+\epsilon
\right\}.
\]
The model predicts first-hit logits $\hat h_{\omega,e}$ over candidate collapsing
edges, trained with a set-valued binary loss:
\[
\mathcal L_{\mathrm{hit}}
=
\mathbb E
\left[
\sum_{e\in E(T_t)}
\mathrm{BCEWithLogits}
\left(
\hat h_{\omega,e},
\mathbf 1[e\in C^\star]
\right)
\right].
\]
An additional false-positive-mass penalty discourages probability mass on edges
outside the first-hit set, which we found necessary to obtain calibrated
set-valued predictions;

\paragraph{Boundary timing loss.}
In addition to classifying the first collapsing edge set, we penalize disagreement
between the predicted velocity-implied hitting times and the supervised boundary
times.  The predicted hitting time is
\[
\hat\tau_e
=
\frac{B_{t,e}}{[-\hat v_{\omega,e}]_+}.
\]
The timing loss encourages the predicted collapse event to occur at the correct
boundary and discourages spurious earlier collapses:
\[
\mathcal L_{\tau}
=
\mathbb E
\left[
\sum_{e\in E(T_t)}
\alpha_e
\left(
\log \hat\tau_e-\log \tau_e^\star
\right)^2
\right],
\]
with larger weight on edges in the first-hit set. On the first-hit set we additionally apply an asymmetric over-prediction penalty
\[
\mathcal L_{\tau}^{\mathrm{over}}
=
\mathbb E\!\left[
\bigl[\,\max_{e\in C^\star}\log\hat\tau_e-\max_{e\in C^\star}\log\tau_e^\star\,\bigr]_+^2
\right],
\qquad [x]_+=\max(x,0),
\]
which discourages first-hit edges from collapsing later than the supervised boundary time but does not penalize earlier collapses; this term is folded into \(\mathcal L_\tau\) with its own weight.

\paragraph{Autoregressive topology loss.}
See Appendix~\ref{Appendix:autoregressive}.

\paragraph{Anchor direct-set auxiliary loss.}
The first-hit head predicts the set of active non-pendant internal split coordinates that collapse at the next BHV orthant boundary. This target is not a function of topology alone: within the same topology, different branch-length vectors can imply different boundary directions. However, along a fixed oracle geodesic phase, the branch lengths vary while the next boundary event remains fixed. We therefore add direct set supervision on precomputed anchor states along each oracle phase.

For each orthant/phase visited by the oracle geodesic, we generate four anchor states by sampling the local BHV geodesic before the next boundary. Each anchor is stored with its phase index, tree state, target tree, oracle velocity labels, and next-boundary tree metadata. During training, we convert each anchor into a direct first-hit target set: the active internal splits that should collapse at the next boundary. The anchor states are passed through the same model as ordinary velocity states, and are supervised with a multi-label first-hit loss on this target set.

\paragraph{Overall objective.}
The reported model is trained with the weighted objective
\[
\mathcal L
=
\lambda_{\mathrm{vel}}\mathcal L_{\mathrm{vel}}
+
\lambda_{\mathrm{hit}}\mathcal L_{\mathrm{hit}}
+
\lambda_{\tau}\mathcal L_{\tau}
+
\lambda_{\mathrm{AR}}\mathcal L_{\mathrm{AR}}
+
\lambda_{\mathrm{anc}}\mathcal L_{\mathrm{anc}}.
\]

This objective should be interpreted as a hybrid path-supervision objective on BHV tree space: continuous flow matching is used within orthants, while topology changes are handled through learned first-hit and autoregressive transition heads rather than through a globally smooth vector field across orthant boundaries. 

\subsection{Autoregressive topology head.}
\label{Appendix:autoregressive}
The autoregressive topology head is invoked only after a boundary collapse has created a polytomy. It does not predict branch lengths. Instead, given the components incident to the polytomy, it predicts which subset of components should be merged in order to introduce the next split.

Let
\[
H=\{h_1,\ldots,h_G\},\qquad h_i\in\mathbb{R}^{128},
\]
denote the variable-size set of component embeddings incident to a polytomy. These embeddings are produced by the graph transformer and therefore already contain information about the current tree topology, branch lengths, sequence conditioning, and discrete phase/time conditioning. Before applying the structured-subset head, each component embedding is further augmented using a learned projection of the component split bitmask, global Phyla context, clade-level Phyla context, and the frozen start-tree code. The start-tree embedding is used additively: the 64-dimensional frozen code is projected to 128 dimensions and added to each component embedding. Given a tree with multiple polytomies this head must choose which polytomy to merge and which components within that polytomy to merge.

With this in mind, it first scores every unordered starter pair $(i,j)$ by applying a two-layer MLP to
\[
[h_i,\ h_j,\ |h_i-h_j|,\ h_i\odot h_j],
\]
producing a starter-pair logit. Second, conditioned on each starter pair, it constructs a pair context and scores every component for membership in the merge subset. This permits merge subsets of size larger than two. Third, a mean-pooled representation of the polytomy is passed through a size head that predicts the merge-subset cardinality; in the final configuration the maximum allowed subset size is 64. A separate scalar \texttt{group\_head} scores which polytomy should be acted on when more than one polytomy is present.

At sampling time, decoding is deterministic. The model selects the highest
scoring starter pair, predicts the subset size, adds the top-scoring additional components under the membership head, takes the union of the selected component masks, and inserts that union as the new split. The newly inserted split is assigned birth length $10^{-3}$.

\paragraph{Autoregressive topology loss.} For a target merge subset $S^\star\subseteq\{1,\ldots,G\}$,
the structured merge loss decomposes as
\[
\mathcal L_{\mathrm{merge}}(S^\star)
=
\mathcal L_{\mathrm{pair}}(S^\star)
+
\mathcal L_{\mathrm{member}}(S^\star)
+
\mathcal L_{\mathrm{size}}(S^\star).
\]

The starter-pair term is a log-softmax loss over unordered component pairs. Because any starter pair fully contained in the target merge subset is valid, we marginalize over all such pairs:
\[
\mathcal L_{\mathrm{pair}}(S^\star)
=
-
\log
\frac{
\sum_{i<j:\,\{i,j\}\subseteq S^\star}
\exp u_{ij}
}{
\sum_{a<b}
\exp u_{ab}
},
\]
where $u_{ij}$ is the starter-pair logit. The membership term is a binary
cross-entropy over component inclusion indicators,
\[
\mathcal L_{\mathrm{member}}(S^\star)
=
\mathrm{BCE}\!\left(
\{\hat y_k\}_{k=1}^G,\,
\{\mathbf 1[k\in S^\star]\}_{k=1}^G
\right),
\]
where the membership logits are conditioned on the starter-pair context. The size term is a cross-entropy loss over the target cardinality:
\[
\mathcal L_{\mathrm{size}}(S^\star)
=
-\log p_{\mathrm{size}}\!\left(|S^\star|\right).
\]

When several target merge subsets are valid for the supervised continuation, the loss is evaluated for each candidate subset and the minimum is used:
\[
\mathcal L_{\mathrm{merge}}
=
\min_{S^\star\in\mathcal A^\star}
\mathcal L_{\mathrm{merge}}(S^\star),
\]
where $\mathcal A^\star$ is the set of valid target merge subsets. If multiple polytomies are present, the model also receives a binary cross-entropy loss from the polytomy-selection head. Thus the total autoregressive loss is
\[
\mathcal L_{\mathrm{AR}}
=
\mathcal L_{\mathrm{merge}}
+
\lambda_{\mathrm{poly}}\mathcal L_{\mathrm{poly}}.
\]
In the final configuration, $\lambda_{\mathrm{poly}}=1.0$. Practically \name almost always predicts binary merges.

\subsection{Frozen start-tree metric encoder}
\label{Appendix:MetricEncoder}

The first-hit and autoregressive heads are conditioned on a per-start-tree
embedding produced by a small encoder that is pretrained before
PhylaFlow training and frozen thereafter. The same encoder weights and the same
embedding table are used throughout PhylaFlow training and at sampling time;
only the embeddings indexed by the current trajectory's start are retrieved.

\paragraph{Architecture (\texttt{SplitSetEncoder}).}
A start tree is represented by its set of internal splits (bipartitions of the
taxa). Each split is encoded as a fixed-width bitmask, projected by a small
MLP, and a global tree embedding is obtained by sum/mean/max pooling over
splits, concatenated with a 2-dimensional taxon-count feature
(normalized count and log-count), and projected to the final embedding
dimension. The output is L2-normalized. Default sizes used for the reported
model: hidden width 256, embedding dimension 64, bitmask width 256.

\paragraph{Training objective.}
The encoder is trained by metric distillation against the normalized
Robinson--Foulds distance $d_{\mathrm{RF}}\in[0,1]$. For a pair of trees
$(T_a,T_b)$ on the same taxon set, with embeddings $u_a,u_b$, the loss is
\[
\mathcal{L}_{\mathrm{enc}}
=
\mathcal{L}_{\mathrm{sim}}
+
w_{\mathrm{dist}}\,\mathcal{L}_{\mathrm{dist}}
+
w_{\mathrm{bin}}\,\mathcal{L}_{\mathrm{bin}}
+
w_{\mathrm{vic}}\,\mathcal{L}_{\mathrm{vic}},
\]
where $\mathcal{L}_{\mathrm{sim}}$ is an MSE between the predicted cosine
similarity $\langle u_a,u_b\rangle$ and the target similarity
$\exp(-d_{\mathrm{RF}}/\tau)$ with $\tau=0.25$;
$\mathcal{L}_{\mathrm{dist}}$ regresses a sigmoid-bounded MLP read-out of
$(u_a,u_b)$ onto $d_{\mathrm{RF}}$; $\mathcal{L}_{\mathrm{bin}}$ is a
cross-entropy loss over discretized $d_{\mathrm{RF}}$ bins
$\{0.05,0.15,0.30,0.50,0.75\}$; and $\mathcal{L}_{\mathrm{vic}}$ is the
VICReg variance/covariance regularizer applied to the batch of embeddings.
Default loss weights: $w_{\mathrm{dist}}=1.0$, $w_{\mathrm{bin}}=0.25$,
$w_{\mathrm{vic}}=0.02$.

\paragraph{Training data.}
Pairs are sampled on the fly from a uniform random tree generator over
unrooted topologies, with three pair types: identical pairs (10\%),
label-swap pairs differing by 1--3 leaf relabelings of the same topology
(40\%), and independent pairs (50\%). All pairs share a common taxon count
drawn from $\{8,16,32,50\}$. Optimizer: AdamW, learning rate $10^{-3}$,
weight decay $10^{-4}$, batch size 64. Training is short by design (a few
hundred to a couple thousand steps); the ``$100$step'' suffix in the released
checkpoint filenames refers to the number of training steps used to fit the
encoder for that artifact.

\paragraph{Embedding-table export.}
After training, the encoder is applied to the dataset-specific start-tree
bank (the same bank used to initialize PhylaFlow trajectories), and the
resulting L2-normalized embeddings are saved as a \texttt{.pt} table indexed
by case ID, alongside metadata (tree hashes, source checkpoint path, and
diagnostic statistics such as inter-embedding cosine similarity and
effective rank). PhylaFlow's first-hit head consumes a 16-dimensional
adapter projection of these embeddings, and the autoregressive head consumes
a 64-dimensional adapter projection (Section~\ref{Appendix: hyperparameters}).

\subsection{Frozen branch-length relaxer}
Topology generation determines the final BHV orthant, but the branch lengths within that orthant remain continuous degrees of freedom. We therefore train a separate topology-preserving branch-length relaxer. Given a fixed topology, the relaxer predicts a within-orthant displacement from an initial branch-length assignment toward the branch lengths obtained after a short fixed-topology MrBayes warmup. The relaxer is not trained to add, remove, or rearrange edges.

For a tree with topology \(T\) and branch lengths \(b=\{b_e\}\), the relaxer predicts an additive update
\[
\Delta_\psi(T,b,E_D)_e
\]
for each represented edge \(e\), conditioned on the target dataset's Phyla (Appendiex~\ref{Appendix:condit:phyla}) embedding bank \(E_D\). Training targets are obtained from topology-frozen MrBayes branch-length warmups. If \(b^{\mathrm{warm}}\) denotes the branch lengths after the fixed-topology warmup initialized from \(b\), the supervised target is
\[
y_e = b^{\mathrm{warm}}_e - b_e .
\]
The relaxer is trained with mean-squared error over matched edges,
\[
\mathcal{L}_{\mathrm{relax}}
=
\frac{1}{|\mathcal{E}(T)|}
\sum_{e\in \mathcal{E}(T)}
\left(
\Delta_\psi(T,b,E_D)_e
-
(b^{\mathrm{warm}}_e-b_e)
\right)^2 .
\]
Edges are matched by their split masks, identifying each split with its complement. The dummy full-mask edge introduced by the rooted encoding is excluded from both the loss and the evaluation-time update.

\paragraph{Relaxer architecture.}
The relaxer reuses a small TreeDenoiserTokenGT graph-transformer trunk to produce topology-aware per-edge features. The trunk uses embedding dimension 64, hidden dimension 64, 2 transformer layers, 4 attention heads, prenorm transformer blocks, exact attention, and zero dropout. Phyla conditioning is supplied through leaf-token additions with 256-dimensional Phyla embeddings. For each edge \(e\), the trunk produces an edge embedding \(z_e\in\mathbb{R}^{64}\). This embedding is concatenated with three scalar edge features: the current branch length, the normalized split size, and an indicator for pendant edges. The delta head is
\[
\mathrm{LayerNorm}
\rightarrow
\mathrm{Linear}(67,128)
\rightarrow
\mathrm{GELU}
\rightarrow
\mathrm{Linear}(128,128)
\rightarrow
\mathrm{GELU}
\rightarrow
\mathrm{Linear}(128,1),
\]
and outputs a raw additive branch-length delta \(\Delta_e\), rather than a log-length delta.

\paragraph{Relaxer training.}
We construct DS1--DS8 branch-relaxation banks by drawing dataset-specific source topologies, perturbing their branch lengths, optionally applying local NNI perturbations to broaden the range of fixed topologies, and then running short topology-frozen MrBayes warmups to obtain target branch lengths. The warmup topology is fixed after these perturbations; the warmup changes only branch lengths. MrBayes uses a JC likelihood model (`nst=1`, equal rates, fixed equal state frequencies), initializes from the input topology and branch lengths, disables topology-changing proposals, and keeps only branch-length movement.

In the final all-dataset relaxer, training uses 5,376 start/warmed tree pairs and 960 heldout pairs across DS1--DS8, with balanced sampling over datasets. The trunk and delta head are optimized jointly with AdamW, batch size 32, learning rate \(2\times 10^{-3}\), weight decay \(10^{-4}\), and gradient clipping at 1.0. The archived checkpoint is selected by heldout relaxed likelihood after applying the predicted branch-length update.

\paragraph{Use at evaluation time.}
After the sampler generates a tree, the relaxer is applied once. For each represented non-dummy edge,
\[
b'_e = \max(b_e + \Delta_\psi(T,b,E_D)_e, 10^{-8}).
\]
The topology \(T\) and its split set are held fixed exactly; only the branch-length coordinates are rewritten. Consequently, topology-only metrics such as Tree-KL, Split-KL, and RF distance are unchanged by the relaxer. The relaxed trees are used only for branch-length-sensitive likelihood evaluation and for downstream MrBayes-start evaluations that require initialized branch lengths.

\subsection{Hyperparameters and Model Settings}
\label{Appendix: hyperparameters}

\begin{table}[h]
\centering
\small
\caption{Architecture hyperparameters of \name.}
\begin{tabular}{ll}
\toprule
\textbf{Component} & \textbf{Value} \\
\midrule
Encoder hidden / embedding dimension & 128 \\
Encoder depth & 4 layers \\
Attention heads & 8 \\
Layer-norm style & prenorm \\
Dropout (attention / activation / drop-path) & 0 \\
Tokenizer Laplacian-PE dimension & 8 \\
Tokenizer encoder depth & 4 \\
Per-taxon sequence embedding dimension (\(\Phi(S)\)) & 256 \\
\midrule
First-hit head case-embedding dimension & 64 \\
First-hit head phase-MLP hidden dim & 256 \\
First-hit head case adapter & 2-layer MLP (hidden 128) \\
Autoregressive head conditioning code dim & 64 \\
Autoregressive head case adapter & 2-layer MLP (hidden 512) \\
\bottomrule
\end{tabular}

\end{table}

\begin{table}[h]
\centering
\small
\caption{Optimizer and loss-weight settings of \name. The
``boundary log-time'' weights correspond to \(\mathcal{L}_\tau\) with a larger
multiplier on edges in the first-hit set; the first-hit-edge weight is
applied to an asymmetric over-prediction term \([\max_{e\in C^\star}\log\hat\tau_e-\max_{e\in C^\star}\log\tau_e^\star]_+^2\) that
penalizes only late collapses (\(\hat\tau>\tau^\star\)) on the first-hit set,
while the all-edges weight is applied to the symmetric squared log-time
error over all active edges.}
\begin{tabular}{ll}
\toprule
\textbf{Setting} & \textbf{Value} \\
\midrule
Optimizer & AdamW, lr \(2\!\times\!10^{-3}\) \\
Gradient clip & 1.0 \\
Velocity / autoregressive optimizer steps & separated \\
Training epochs & up to 200 (best checkpoint by validation) \\
Batch size & 1 path \\
Warmup steps & 0 \\
Random seed & 42 \\
\midrule
\(\lambda_{\mathrm{vel}}\) (continuous velocity MSE) & 1.0 \\
\(\lambda_{\mathrm{AR}}\) (autoregressive cross-entropy) & 0.2 \\
First-hit BCE weight & 10.0 \\
Boundary log-time weight (all edges) & 2.0 \\
Boundary log-time weight (first-hit edges, over) & 10.0 \\
Event precision margin / weight & \(1.0\,/\,10.0\) \\
Probe direct-set anchor loss & enabled (target-negative weight 1.0) \\
\bottomrule
\end{tabular}
\end{table}

\subsection{Deterministic event-based sampling}
\label{Appendix:Sampling}

We generate trees with a deterministic event-based rollout in BHV tree space. The sampler takes as input an initial tree \(T^{(0)}\), precomputed Phyla sequence embeddings \(\Phi(S)\) for the dataset, a start-tree case embedding \(c\), and a trained PhylaFlow checkpoint with parameters \(\omega\). We write the tree at discrete phase \(k\) as
\[
T^{(k)} = \left(\mathcal{S}^{(k)}, \ell^{(k)}\right),
\]
where \(\mathcal{S}^{(k)}\) denotes the topology/split set and
\(\ell^{(k)}\) denotes branch lengths. The reported experiments use at most \(K_{\max}=8\) discrete phases, at most \(R_{\max}=128\) continuous rollout steps, and at most \(M_{\max}=500\) autoregressive split-insertion events.

\paragraph{Model evaluation at a phase.}
At phase \(k\), the current tree state is tokenized together with the
sequence-derived embeddings, the continuous within-orthant time \(\tau\), the discrete phase index \(k\), and the start-tree case embedding \(c\). The resulting graph, node, and edge tokens are passed through the graph transformer, producing hidden edge representations \(h_e\). The velocity head predicts a scalar branch-length velocity for each represented edge,
\[
\hat v_{\omega,e} = g_{\mathrm{vel}}(h_e),
\qquad e \in E(T^{(k)}),
\]
while the first-hit head predicts a logit for each active internal edge,
\[
\hat h_{\omega,e}
=
g_{\mathrm{hit}}(h_e,\tau,k;c),
\qquad e \in E_{\mathrm{int}}(T^{(k)}).
\]
Pendant edges are included in the representation and may receive velocity
predictions, but they are not treated as topology-changing coordinates.

\paragraph{Decoding the boundary-collapse set.}
Let
\[
C_k
=
\left\{
e \in E_{\mathrm{int}}(T^{(k)}) :
\ell^{(k)}_e > \epsilon
\right\},
\qquad
\epsilon = 10^{-8},
\]
be the set of active internal edges. If \(C_k\) is empty, the rollout
terminates. Otherwise, the first-hit logits are decoded deterministically. We
first form the positive-logit set
\[
F_k^+
=
\left\{
e \in C_k : \hat h_{\omega,e} > 0
\right\}.
\]
The predicted collapse set is then
\[
F_k =
\begin{cases}
F_k^+, & \text{if } F_k^+ \neq \emptyset,\\[4pt]
\left\{\arg\max_{e\in C_k} \hat h_{\omega,e}\right\},
& \text{otherwise.}
\end{cases}
\]
The sampler proceeds only if at least one selected edge has negative predicted
velocity. Define
\[
F_k^-
=
\left\{
e \in F_k : \hat v_{\omega,e} < 0
\right\}.
\]
If \(F_k^-=\emptyset\), the rollout terminates.

\paragraph{Exact-boundary branch-length update.}
We use exact-boundary mode rather than a fixed step size. The boundary-collapse step size is
\[
\Delta \tau_k
=
\max_{e\in F_k^-}
\frac{\ell^{(k)}_e}{-\hat v_{\omega,e}}.
\]
Branch lengths are first updated according to the predicted velocity field,
\[
\widetilde{\ell}_e
=
\ell^{(k)}_e
+
\Delta \tau_k \hat v_{\omega,e}.
\]
All edges in the predicted collapse set \(F_k\) are then clamped exactly to
zero. Active internal edges that were not selected for collapse are floored to
remain positive:
\[
\ell'_e =
\begin{cases}
0, & e \in F_k,\\[4pt]
\max\{\widetilde{\ell}_e,\epsilon\},
& e \in E_{\mathrm{int}}(T^{(k)}) \setminus F_k.
\end{cases}
\]
The tree is then reconstructed from the remaining positive internal split
coordinates, together with the updated pendant branch lengths. Collapsing one
or more internal edges may create an unresolved polytomy.

\paragraph{Autoregressive topology resolution.}
If the boundary collapse creates a polytomy, we resolve it with the topology head. The current polytomous tree is passed back through the same graph transformer at phase \(k\), using the same Phyla sequence conditioning \(\Phi(S)\), continuous time input, and start-tree conditioning \(c\). For each polytomy, the structured-subset topology head scores candidate component merges. This head predicts a starter pair, a subset size, and component membership scores. Decoding selects the highest-scoring valid subset of polytomy components.

The union of the selected components defines a candidate new split. If the
split is valid and is not already present in the tree, it is inserted with fixed birth length
\[
\ell_{\mathrm{birth}} = 10^{-3},
\]
and the tree is rebuilt. By default, we apply only one planned autoregressive merge per topology-head forward pass. We then repeat topology decoding until the current tree has no remaining polytomy, no valid merge is available, or the autoregressive event budget is exhausted.

\paragraph{Stopping criteria and generated distribution.}
After the boundary collapse and any autoregressive topology resolution, the phase index is incremented, \(k \leftarrow k+1\). The rollout terminates when the maximum number of phases is reached, the rollout-step budget is exhausted, the autoregressive event budget is exhausted, no selected first-hit edge has negative velocity, or no valid topology merge can be found. The final generated tree is the output of this deterministic rollout. After this we then run the branch-length relaxer and return the tree.

\section{Additional experimental details}

\subsection{Benchmark datasets and posterior references}

We evaluate on the DS1--DS8 benchmark datasets used in prior phylogenetic posterior-inference studies \citep{lakner2008mcmc}. For each dataset, we use two \textsc{MrBayes} posterior runs. A short run is used to construct training targets for \name, while a longer run is used as the reference posterior for evaluation. The long-run posterior defines the empirical topology distribution $\widehat{\pi}^{\mathrm{long}}_d$ used when computing TreeKL.

For each dataset, we extract the unique topologies observed in the short-run posterior and their empirical frequencies. Training pairs are then constructed by sampling target posterior topologies in proportion to their empirical frequency and pairing each target with a random start tree. We compute the BHV geodesic between each start tree and target tree and train \name on these start-to-posterior paths.

At evaluation time, \name is applied to unseen random start trees. The number of generated start trees, and hence the number of downstream \textsc{MrBayes} chains, is dataset-specific based on the number of unique topologies in the posterior, for DS1 this was 78, DS2: 42, DS3: 243, DS4: 573, DS5: 525, DS6: 219, DS7: 1344, and DS8: 1122. For each dataset, every initialization method is evaluated using the same number of starting trees and the same downstream \textsc{MrBayes} configuration, except for PhyloGFN where we used 1000 trees for each dataset due to the recommendation of the paper. We found sub-sampling PhyloGFN trees did not substantially change results.

\subsection{Random start tree generation}
\label{Appendix:start_tree_generation}

A random start tree is generated on the dataset taxon set by sequential random edge insertion. We first sort the taxa by their dataset labels and initialize the tree with the unique fully resolved unrooted topology on the first three taxa. For each remaining taxon in sorted order, we choose one current edge uniformly at random, subdivide that edge with a new internal node, and attach the new taxon to that node. This produces a fully resolved unrooted topology. Branch lengths for all non-artificial-root edges are sampled independently from the continuous uniform distribution on $[0.1,1.0]$. 

\subsection{Metrics}
\label{Appendix: Metrics}

\paragraph{Tree-KL} Our primary topology-distribution metric is a smoothed TreeKL divergence between the long-run reference posterior topology distribution and the topology distribution induced by a finite-budget chain. Let $c_{\mathrm{ref}}(x)$ and $c_{\mathrm{sample}}(x)$ denote the number of times topology $x$ appears in the reference and finite-budget samples, respectively, with totals $N_{\mathrm{ref}}$ and $N_{\mathrm{sample}}$. We compare the two empirical distributions on the union of their observed supports,
\[
\mathcal{A}
=
\mathrm{supp}(c_{\mathrm{ref}})
\cup
\mathrm{supp}(c_{\mathrm{sample}}),
\]
using additive smoothing with $\alpha=10^{-6}$:
\[
p_{\alpha}(x)
=
\frac{c_{\mathrm{ref}}(x)+\alpha}
{N_{\mathrm{ref}}+\alpha|\mathcal{A}|},
\qquad
q_{\alpha}(x)
=
\frac{c_{\mathrm{sample}}(x)+\alpha}
{N_{\mathrm{sample}}+\alpha|\mathcal{A}|}.
\]
The reported TreeKL is the reference-to-sample KL divergence
\[
\mathrm{TreeKL}
=
\mathrm{KL}(p_{\alpha}\Vert q_{\alpha})
=
\sum_{x\in\mathcal{A}}
p_{\alpha}(x)
\log
\frac{p_{\alpha}(x)}{q_{\alpha}(x)}.
\]
Thus, TreeKL measures how well the finite-budget sample covers the topology mass of the long-run reference posterior. Topologies present in the reference but absent from the finite-budget sample are assigned small but nonzero smoothed sample probability, so their contribution is large but finite. Lower TreeKL indicates closer agreement with the reference posterior topology distribution.

\paragraph{Split-KL}
TreeKL compares distributions over exact topologies. As a coarser diagnostic, we also report SplitKL, which compares the induced distributions over nontrivial bipartitions of the leaf set. For leaf set $L$, a split is an unordered bipartition of $L$ of the form
$B \mid L\setminus B$, where $B\subset L$ is a subset of leaves.
We call a split nontrivial if both sides contain at least two leaves, i.e.,
\[
2 \le |B| \le |L|-2,
\]
and identify $B \mid L\setminus B$ with its complement$(L\setminus B)\mid B$. For an unrooted topology $x$, let $\mathcal{S}(x)$ denote the set of nontrivial splits induced by deleting internal edges of $x$, with each split represented in this canonical complement-invariant form. For a fully resolved unrooted tree with $n$ leaves, $|\mathcal{S}(x)|=n-3$.

For a collection of sampled topologies, define split counts by accumulating split occurrences over trees:
\[
c^{\mathrm{split}}_{\mathrm{ref}}(s)
=
\sum_x c_{\mathrm{ref}}(x)\mathbf{1}\{s\in\mathcal{S}(x)\},
\qquad
c^{\mathrm{split}}_{\mathrm{sample}}(s)
=
\sum_x c_{\mathrm{sample}}(x)\mathbf{1}\{s\in\mathcal{S}(x)\}.
\]
Let
\[
M_{\mathrm{ref}}
=
\sum_s c^{\mathrm{split}}_{\mathrm{ref}}(s),
\qquad
M_{\mathrm{sample}}
=
\sum_s c^{\mathrm{split}}_{\mathrm{sample}}(s),
\]
and compare the two split distributions on the union of observed nontrivial split supports,
\[
\mathcal{A}_{\mathrm{split}}
=
\mathrm{supp}(c^{\mathrm{split}}_{\mathrm{ref}})
\cup
\mathrm{supp}(c^{\mathrm{split}}_{\mathrm{sample}}).
\]
Using the same additive smoothing parameter $\alpha=10^{-6}$, define
\[
p^{\mathrm{split}}_{\alpha}(s)
=
\frac{c^{\mathrm{split}}_{\mathrm{ref}}(s)+\alpha}
{M_{\mathrm{ref}}+\alpha|\mathcal{A}_{\mathrm{split}}|},
\qquad
q^{\mathrm{split}}_{\alpha}(s)
=
\frac{c^{\mathrm{split}}_{\mathrm{sample}}(s)+\alpha}
{M_{\mathrm{sample}}+\alpha|\mathcal{A}_{\mathrm{split}}|}.
\]
The reported SplitKL is
\[
\mathrm{SplitKL}
=
\mathrm{KL}
\!\left(
p^{\mathrm{split}}_{\alpha}
\Vert
q^{\mathrm{split}}_{\alpha}
\right)
=
\sum_{s\in\mathcal{A}_{\mathrm{split}}}
p^{\mathrm{split}}_{\alpha}(s)
\log
\frac{
p^{\mathrm{split}}_{\alpha}(s)
}{
q^{\mathrm{split}}_{\alpha}(s)
}.
\]
For fully resolved unrooted trees with $n$ leaves, each topology contributes $n-3$ nontrivial splits, so before smoothing this count-based split distribution is equivalent to the posterior split marginal normalized by $n-3$. Lower SplitKL indicates closer agreement with the reference posterior at the level of individual posterior splits. Unlike TreeKL, SplitKL ignores correlations among splits and therefore does not require exact topology matches; it measures whether the sampled trees recover the reference posterior split mass.

\label{Appendix:likelihood_diagnostic}
\paragraph{Likelihood} We evaluate all methods under a common fixed-topology optimized-likelihood diagnostic. For each candidate topology \(T\) and alignment \(X\), we run IQ-TREE \citep{IQTree} with topology fixed and score
  \[
  \ell_{\mathrm{opt}}(T)
  =
  \max_{\mathbf{b}}
  \log p\!\left(X \mid T, \mathbf{b}, Q_{\mathrm{F81}}(\hat{\pi})\right),
  \]
where \(\mathbf{b}\) are branch lengths optimized by IQ-TREE, \(Q_{\mathrm{F81}}\) is the F81 substitution model, \(\hat{\pi}\) are empirical nucleotide frequencies from the alignment (IQ-TREE F81+F), and among-site rates are uniform. We fix the topology, optimize branch lengths, and use no gamma-rate heterogeneity, invariant-sites component, topology prior, or branch-length prior.

Because the provided posterior for the DS1-DS8 runs had no .p files from the \textsc{MrBayes} run we had to approximate the likelihood of the posterior trees. The posterior-likelihood reference is built from the highest posterior-mass unique topologies in the long-chain reference .trprobs files. We score these reference topologies with the same fixed-topology `F81+F` procedure and use their posterior-weighted median optimized log-likelihood as the reference location. 

We then define the gap to the reference median log-likelihood as:

  \[
  \Delta_{\mathrm{opt}} =
  \operatorname{median}_{\mathrm{ref}}(\ell_{\mathrm{opt}})
  -
  \operatorname{median}_{\mathrm{method}}(\ell_{\mathrm{opt}}).
  \]

Smaller values indicate that the method's topology distribution lies closer to the posterior-core optimized-likelihood basin; negative values indicate that the method median is above the reference median. We chose representation topologies from the posterior by sampling topologies from the final portion of the \textsc{MrBayes} chains, retaining samples with generation at least 80K and subsampling up to 150 topologies.

\subsection{MrBayes evaluation protocol}

All initialization methods are evaluated by running \textsc{MrBayes} from method-specific starting trees. Unless otherwise stated, all runs use:
\begin{itemize}
    \item \texttt{ngen=100000},
    \item \texttt{samplefreq=200},
    \item one independent chain per starting tree,
    \item cumulative pooling of samples across chains.
\end{itemize}

For each method, dataset, and checkpoint, we pool all sampled topologies from generation 0 up to the checkpoint and compute the empirical topology distribution. We report TreeKL against the long-run reference posterior at 20K, 40K, 60K, 80K, and 100K generations. We also compute the average log likelihood of sampled states at each checkpoint.

\subsection{Estimating the runtime of PhyloGFN and short-runs}
\label{Appendix:run_time_estimate}

\begin{table}[t]
\centering
\small
\setlength{\tabcolsep}{5.0pt}
\renewcommand{\arraystretch}{1.08}
\caption{\textbf{Pre-refinement setup time for strong controls.}
Times reported in hours. For short-warmup, we report the lower-bound runtime. For \name, we report a conservative setup cost equal to \name training time plus the optimistic short-run supervision cost used to construct training targets (Appendix~\ref{Appendix:run_time_estimate}).}
\label{tab:setup_runtime_strong_controls}
\begin{tabular}{lrrrrrrrr}
\toprule
Method & DS1 & DS2 & DS3 & DS4 & DS5 & DS6 & DS7 & DS8 \\
\midrule
short-warmup 
& 0.72 & 0.77 & 0.95 & 1.05 & 0.47 & 0.56 & 1.00 & 0.30 \\
PhyloGFN 
& 15.67 & 17.32 & 20.08 & 25.97 & 31.97 & 33.80 & 43.52 & 47.60 \\
\midrule
\name 
& 18.73 & 12.73 & 13.63 & 9.95 & 25.87 & 19.02 & 14.68 & 25.62 \\
\bottomrule
\end{tabular}
\end{table}

\paragraph{PhyloGFN training runtime estimate.}
We estimate the wall-clock training time for the 24\% PhyloGFN setting using theruntime measurements reported in Appendix~G of the PhyloGFN ICLR paper~\citep{zhou2024phylogfn}. The paper reports full-training runtimes for DS1--DS8 in Table~S6 and directly reports the DS1 24\% runtime in Table~S7. For DS1, we therefore use the reported 24\% runtime directly. For DS2--DS8, where 24\% runtimes are not separately reported, we extrapolate from the full runtime using the empirical DS1 scaling factor:
\[
\alpha
=
\frac{T_{\text{DS1},\,24\%}}{T_{\text{DS1},\,\mathrm{Full}}}
=
\frac{15\mathrm{h}\,40\mathrm{m}}{62\mathrm{h}\,40\mathrm{m}}
=
0.25.
\]
Thus, for each dataset $d \in \{\mathrm{DS2},\ldots,\mathrm{DS8}\}$, we estimate
\[
\widehat{T}_{d,\,24\%}
=
0.25 \cdot T_{d,\,\mathrm{Full}}.
\]

These estimates are intended only as training wall-clock estimates under the hardware and implementation settings reported by the PhyloGFN paper (Table ~\ref{tab:phylogfn_runtime_estimate}). They do not include additional time for posterior sample generation, our 1000-sample evaluation procedure, or downstream MrBayes refinement.

\paragraph{Short-run runtime estimate }
\label{app:mrbayes-short-run-runtime}

The MrBayes short-run reference procedure used in our comparison consists of 10 independent MrBayes runs, each run until the average standard deviation of split frequencies (ASDSF) is below 0.01 or until a maximum of 100 million iterations is reached. Tree topologies are sampled every 100 iterations, and the first 25\% of iterations are discarded as burn-in. The exact stopping iteration and hardware used for the original short-run references were not available. Therefore, we estimate the short-run runtime by scaling same-hardware measurements of one cold MrBayes chain run for 1 million iterations. Let \(T_{1M,d}\) denote the wall-clock time for one cold 1-million-iterationMrBayes chain on dataset \(d\). 

We report two estimates. The optimistic estimate assumes that each of the 10 independent MrBayes runs converges by 10 million iterations, that the 10 runs are executed in parallel, that runtime scales linearly with iteration count, and that scheduling, sampling, I/O, and post-processing overheads are negligible. Under these assumptions, the wall-clock time for dataset \(d\) is $10*T_{1M,d}$ with corresponding total compute of $100*T_{1M,d}$ because the protocol runs 10 independent chains. The pessimistic estimate assumes that the runs reach the 100-million-iteration cap, but still assumes that the 10 independent runs are parallelized. Thus, the pessimistic estimate is $1000*T_{1M,d}$ (Table~\ref{tab:mrbayes-short-run-walltimes}). 

\begin{table}[t]
\centering
\small
\begin{tabular}{lccc}
\toprule
Dataset &
One 1M chain time &
Optimistic wall time &
Cap-case wall time \\
& & 10M, 10 runs parallelized & 100M, 10 runs parallelized \\
\midrule
DS1 & 4.3--4.6 min & 43--46 min & 7.2--7.7 h \\
DS2 & 4.6--4.9 min & 46--49 min & 7.7--8.2 h \\
DS3 & 5.7 min & 57 min & 9.5 h \\
DS4 & 6.3--6.7 min & 63--67 min & 10.5--11.2 h \\
DS5 & 2.8--4.1 min & 28--41 min & 4.7--6.8 h \\
DS6 & 3.36--3.40 min & 33.6--34.0 min & 5.6--5.7 h \\
DS7 & 6--9 min & 60--90 min & 10--15 h \\
DS8 & 1.89--2.04 min & 18.9--20.4 min & 3.15--3.40 h \\
\bottomrule
\end{tabular}
\caption{
Estimated wall-clock runtimes for the MrBayes short-run reference chains under
two stopping assumptions. The optimistic estimate assumes convergence by 10M
iterations. The cap-case estimate assumes that the runs reach the 100M
iteration cap. Both estimates assume that the 10 independent MrBayes runs are
parallelized.
}
\label{tab:mrbayes-short-run-walltimes}
\end{table}

\begin{table}[t]
\centering
\caption{
Estimated PhyloGFN 24\% training runtimes. DS1 is the runtime directly reported
in Table~S7 of the PhyloGFN ICLR paper; DS2--DS8 are extrapolated as
$0.25\times$ the full-training runtimes reported in Table~S6.
}
\label{tab:phylogfn_runtime_estimate}
\begin{tabular}{lrrl}
\toprule
Dataset & Full training runtime & 24\% training runtime & Basis \\
\midrule
DS1 & 62h 40m  & 15h 40m & Reported \\
DS2 & 69h 16m  & 17h 19m & $0.25\times$ Full \\
DS3 & 80h 20m  & 20h 05m & $0.25\times$ Full \\
DS4 & 103h 52m & 25h 58m & $0.25\times$ Full \\
DS5 & 127h 52m & 31h 58m & $0.25\times$ Full \\
DS6 & 135h 12m & 33h 48m & $0.25\times$ Full \\
DS7 & 174h 04m & 43h 31m & $0.25\times$ Full \\
DS8 & 190h 24m & 47h 36m & $0.25\times$ Full \\
\bottomrule
\end{tabular}
\end{table}

\subsection{Initialization methods}
\label{Appendix:classic_methods}

\paragraph{\name.}
For each dataset, we train \name using BHV geodesic paths from random start trees to topologies sampled from the short-run posterior distribution. At evaluation time, we apply the trained model to unseen random start trees and use the resulting trees to initialize \textsc{MrBayes}. These evaluation starts are not present in the training bank.

\paragraph{Random initialization.}
As a baseline, we initialize \textsc{MrBayes} from random trees sampled independently from the same start-tree distribution used for \name evaluation.

\paragraph{IQ-TREE maximum likelihood.}
We compute a maximum-likelihood tree using \textsc{IQ-TREE} with automatic model selection:
\begin{verbatim}
iqtree3 -s alignment.nex -m MFP --runs 10 -T 4 --seed ...
\end{verbatim}
This produces a single ML topology. For datasets requiring multiple chains, the same ML topology is reused as the starting topology across chains, with independent \textsc{MrBayes} trajectories thereafter.

\paragraph{MrBayes parsimony.}
We generate parsimony-based starting trees using \textsc{MrBayes}' built-in parsimony initialization:
\begin{verbatim}
starttree=parsimony nperts=0
\end{verbatim}
For each chain, we extract the generation-0 parsimony tree and use it as the starting tree for the corresponding 100K-generation run.

\paragraph{UShER + matOptimize.}
We construct parsimony-refined start trees using a UShER/matOptimize pipeline \citep{turakhia_ultrafast_2021}. The alignment is converted into UShER-compatible input, an initial tree is constructed from a closest-pair cherry, taxa are added by greedy placement with \textsc{UShER}, and the resulting mutation-annotated tree is refined with \textsc{matOptimize}. In the single-start variant, the optimized topology is reused across all chains.

\paragraph{UShER + matOptimize multistart.}
We also evaluate a multistart version of the UShER/matOptimize pipeline. Instead of using a single initial cherry, we sample multiple initial taxon pairs and run the full greedy-placement and matOptimize refinement procedure from each. This produces a diverse collection of parsimony-refined classical starting trees, which are then used to initialize separate \textsc{MrBayes} chains.

\subsection{Tree-KL Extended Results}
\label{Appendix:tree_kl_extended_results}

\paragraph{Longer \textsc{MrBayes} Chain Sensitivity Analysis}
\label{app:mrbayes-long-chain}

To test whether the $10^5$-generation \textsc{MrBayes} runs used in the main experiments were
artificially short, we repeated the fixed-start \textsc{MrBayes} evaluation with substantially
longer chains. For each dataset and initialization condition, we reran the same fixed-start
\textsc{MrBayes} harness used in the main $10^5$-generation comparison, changing only the chain
length to $10^6$ generations. Each start tree was run as an independent cold \textsc{MrBayes} chain
with \texttt{nruns=1}, \texttt{nchains=1}, \texttt{starttree=current}, and \texttt{nperts=0}. We
sampled every 200 generations and recomputed the pooled topological posterior after 100K, 250K,
500K, and 1M generations by aggregating samples across all independent chains for that condition.
Thus, the intermediate columns in Table~\ref{tab:mrbayes-long-chain-sensitivity} are prefixes of
the same $10^6$-generation runs, not separate shorter reruns.

We evaluated two initialization conditions: the PhylaFlow-selected start set used in the final
results, and the strongest non-PhylaFlow \textsc{MrBayes}-100K baseline for that dataset. The
non-PhylaFlow baseline was chosen separately per dataset from the best 100K endpoint among the
classical and random initializations. Runtime was measured as wall-clock time for the full block
using up to 10 parallel workers; it is therefore not the runtime of a single chain. Lower TreeKL is
better.

\begin{table}[t]
\centering
\small
\setlength{\tabcolsep}{4pt}
\caption{Sensitivity of fixed-start \textsc{MrBayes} results to extending chains from 100K to 1M
generations. TreeKL is computed against the golden posterior over tree topologies. Each row pools
samples from all independent chains in that condition.}
\label{tab:mrbayes-long-chain-sensitivity}
\begin{tabular}{llrrrrrrl}
\toprule
Dataset & Initialization & 100K & 250K & 500K & 1M \\
\midrule
DS1 & PhylaFlow & 0.143 & 0.114 & 0.100 & 0.088 \\
DS1 & Best non-PF & 0.405 & 0.207 & 0.140 & 0.110 \\
\midrule
DS2 & PhylaFlow & 0.054 & 0.054 & 0.055 & 0.055 \\
DS2 & Best non-PF  & 0.061 & 0.056 & 0.052 & 0.052 \\
\midrule
DS3 & PhylaFlow & 2.343 & 2.451 & 2.457 & 2.427 \\
DS3 & Best non-PF  & 2.079 & 2.279 & 2.324 & 2.365 \\
\midrule
DS4 & PhylaFlow & 0.174 & 0.164 & 0.164 & 0.163 \\
DS4 & Best non-PF & 0.188 & 0.171 & 0.164 & 0.163\\
\midrule
DS5 & PhylaFlow & 3.508 & 3.461 & 3.441 & 3.397 \\
DS5 & Best non-PF  & 4.049 & 3.792 & 3.631 & 3.509\\
\midrule
DS6 & PhylaFlow & 4.899 & 4.970 & 4.992 & 4.993 \\
DS6 & Best non-PF  & 5.002 & 5.012 & 5.009 & 5.006 \\
\bottomrule
\end{tabular}
\end{table}

The long-chain sensitivity analysis supports two conclusions (Table~\ref{tab:mrbayes-long-chain-sensitivity}). First, some standard
\textsc{MrBayes} baselines do improve with longer chains, as expected; for example, the best
non-PhylaFlow DS1 baseline improves from TreeKL $0.405$ at 100K generations to $0.110$ at 1M
generations. Second, this improvement does not explain away the main hard-case gains from
PhylaFlow-guided refinement. On DS3, the best completed non-PhylaFlow fixed-start baseline reaches
TreeKL $2.365$ after 1M generations, while PhylaFlow-MCMC reaches $0.117$ under the main
finite-budget setting. On DS6, the best completed non-PhylaFlow fixed-start baseline reaches
TreeKL $5.006$ after 1M generations, while PhylaFlow-MCMC reaches $4.479$. Therefore, the observed
benefits of PhylaFlow-MCMC are not solely attributable to the standard \textsc{MrBayes} baselines
being truncated at 100K generations.

\begin{table}[ht]
\centering
\small
\setlength{\tabcolsep}{4.5pt}
\renewcommand{\arraystretch}{1.08}
\caption{Best Tree-KL attained during the finite-budget run, relative to the long-run posterior topology distribution. Smaller values indicate closer agreement with the posterior topology distribution. Best values per dataset are bolded.}
\label{tab:tree_kl_best_supplement}
\begin{tabular}{lrrrrrrrr}
\toprule
Method & DS1 & DS2 & DS3 & DS4 & DS5 & DS6 & DS7 & DS8 \\
\midrule
Random       & 0.403 & 0.072 & 1.945 & 0.263 & 4.274 & 5.097 & 2.620 & 0.905 \\
MP           & 0.430 & 0.071 & 1.106 & 0.277 & 4.215 & 4.951 & 2.217 & 0.987 \\
IQ-TREE      & 3.699 & 0.060 & 1.689 & 0.184 & 6.790 & 5.255 & 2.651 & 0.989 \\
UShER-s      & 0.586 & 0.084 & 2.476 & 0.294 & 4.823 & 5.063 & 2.676 & 0.939 \\
UShER-ms     & 0.435 & 0.082 & 2.489 & 0.273 & 4.136 & 5.057 & 2.451 & 0.821 \\
PhylaFlow    & \textbf{0.150} & \textbf{0.049} & 1.081 & \textbf{0.169} & 3.264 & 4.814 & 2.468 & 0.787 \\
PhylaFlow-MCMC & 0.304 & 0.139 & \textbf{0.117} & 0.514 & \textbf{1.659} & \textbf{4.432} & \textbf{1.071} & \textbf{0.612} \\
\bottomrule
\end{tabular}
\end{table}

\subsection{Likelihood Results}

\begin{table}[ht]
\centering
\small
\setlength{\tabcolsep}{4.5pt}
\renewcommand{\arraystretch}{1.08}
\caption{Optimized log-likelihood shortfall relative to the posterior-core reference median. Panel A evaluates the initial topology before any MrBayes updates. Panel B evaluates median tail-sampled topologies after finite-budget MCMC. Smaller gaps indicate closer agreement with the posterior-core optimized-likelihood basin; negative values indicate optimized likelihoods above the reference median.}
\label{tab:init_tail_likelihood_gap}
\begin{tabular}{lrrrrrrrr}
\toprule
Method & DS1 & DS2 & DS3 & DS4 & DS5 & DS6 & DS7 & DS8 \\
\midrule
\multicolumn{9}{l}{\textbf{Initial topology, before MrBayes}} \\
\midrule
Random   & 2316.633 & 5218.048 & 7945.339 & 3847.616 & 2324.115 & 4423.504 & 21250.269 & 5293.962 \\
MP       & 171.299  & 183.554  & 44.832   & 353.745  & 96.568   & 99.920   & 55.334    & 222.996 \\
IQ-TREE  & \textbf{-1.429} & 38.782 & 35.054 & 23.979 & 22.324 & \textbf{3.452} & \textbf{40.349} & 13.031 \\
UShER-s  & 53.801   & 372.558  & 977.494  & 3475.298 & 245.546  & 75.432   & 777.055   & 178.183 \\
UShER-ms & 161.952  & 389.393  & 622.486  & 9233.202 & 213.853  & 148.411  & 855.326   & 150.154 \\
\midrule
PhylaFlow & 11.011 & \textbf{2.395} & \textbf{-1.627} & \textbf{-1.265} & \textbf{-0.580} & 14.932 & 164.745 & \textbf{-1.367} \\
\midrule
\multicolumn{9}{l}{\textbf{100K-step tail samples, after finite-budget MCMC}} \\
\midrule
Random   & 0.757 & \textbf{0.000} & \textbf{1.061} & 1.183 & 10.286 & 2.414 & 2.772 & 1.584 \\
MP       & 0.692 & \textbf{0.000} & \textbf{1.061} & \textbf{0.000} & 9.427 & 2.086 & 2.772 & 1.585 \\
IQ-TREE  & \textbf{-0.925} & \textbf{0.000} & \textbf{1.061} & \textbf{0.000} & 10.978 & 2.084 & 2.772 & 1.200 \\
UShER-s  & 0.692 & \textbf{0.000} & \textbf{1.061} & \textbf{0.000} & 9.349 & \textbf{1.690} & \textbf{1.728} & 1.704 \\
UShER-ms & 0.692 & \textbf{0.000} & \textbf{1.061} & \textbf{0.000} & 10.169 & 2.086 & 2.770 & \textbf{1.009} \\
\midrule
PhylaFlow & 0.555 & \textbf{0.000} & \textbf{1.061} & \textbf{0.000} & \textbf{8.246} & 2.766 & 1.801 & 2.950 \\
\bottomrule
\end{tabular}
\end{table}

\subsection{PhylaFlow-MCMC details}
\label{Appendix:phylaflowmcmc}

Our fixed-budget results suggest that learned initialization and posterior refinement solve related but distinct problems. A PhylaFlow start bank can place \textsc{MrBayes} close to high-posterior regions of tree space, improving the finite-budget tree distribution in many datasets. However, initialization alone is passive: after the start trees are handed to \textsc{MrBayes}, the subsequent trajectory is controlled by the standard proposal kernel. On difficult datasets, the chain may still spend much of its budget in an incomplete or suboptimal posterior basin, even when the initial trees contain useful posterior signal. This motivates a second use of the learned model: not as a replacement for Bayesian sampling, but as a proposal guide that biases local moves toward tree topologies whose splits are supported by the learned PhylaFlow distribution. The goal of PhylaFlow-MCMC is therefore to increase finite-budget posterior basin occupancy: enter high-posterior topology regions earlier and remain there for more of the run.

Let \(x=(\tau,b)\) denote a phylogenetic state consisting of a topology \(\tau\) and branch lengths \(b\), and let
\[
\pi(x) \propto p(Y\mid x)p(x)
\]
be the target posterior density under the same likelihood model used by the evaluation runs. PhylaFlow first produces a bank of guide trees \(G=\{\tau_i\}\). From this guide bank we estimate split weights
\[
\widehat p(s) = \max\left\{\epsilon,\frac{c_G(s)}{\sum_{s'}c_G(s')}\right\},
\]
where \(c_G(s)\) is the count of split \(s\) in the guide bank and \(\epsilon\) is a floor. For a candidate topology \(\tau\), define the guide score
\[
S(\tau)=\sum_{s\in \mathcal S(\tau)} \log \widehat p(s).
\]
At each sweep over the current chain states, a chain attempts a proposal with probability \(p_{\mathrm{prop}}\). If a proposal is attempted, we enumerate local NNI neighbors \(\mathcal N(x)\), optionally relax branch lengths locally after the NNI move, and sample a candidate \(x'\in\mathcal N(x)\) from a mixed proposal
\[
q(x'\mid x)
=
(1-\rho)\frac{1}{|\mathcal N(x)|}
+
\rho
\frac{\exp\{\lambda[S(\tau')-S(\tau)]\}}
{\sum_{z\in\mathcal N(x)}\exp\{\lambda[S(\tau_z)-S(\tau)]\}}.
\]
Here \(\lambda\) controls the strength of split guidance and \(\rho\) mixes guided and uniform local proposals. We then apply the Metropolis-Hastings correction
\[
\alpha(x,x')
=
\min\left\{
1,
\frac{\pi(x')q(x\mid x')}{\pi(x)q(x'\mid x)}
\right\},
\]
where the reverse proposal probability is computed over the reverse NNI neighborhood. Thus the learned model affects proposal efficiency, while the acceptance decision remains likelihood- and proposal-corrected. In the reported runs we use the branch-state split-guided mixed NNI kernel with \(p_{\mathrm{prop}}=0.05\), \(\rho=0.1\), \(\lambda=0.5\), BHV split scoring, and local branch warming/relaxation.

\subsection{PhylaFlow-MCMC versus PhyloGFN-MCMC}
\label{Appendix:PhyloGFNMCMC}
To assess whether the advantage of PhylaFlow-MCMC on D3, DS5, DS6, and DS7 was due to the MCMC procedure itself rather than the PhylaFlow initialization, we ran a branch-warmed variant of PhyloGFN, which we call PhyloGFN-MCMC. This variant used the same MCMC procedure as PhylaFlow-MCMC, but with initial trees and splits provided by PhyloGFN. As shown in Table~\ref{tab:phylogfn_phylaflow_mcmc_treekl}, PhyloGFN-MCMC achieved lower tree-KL than PhylaFlow-MCMC on DS3, but PhylaFlow-MCMC achieved substantially lower tree-KL on the harder posterior distributions DS5, DS6, and DS7.

\begin{table}[t]

\centering
\caption{Tree-KL comparison for branch-warmed MCMC runs. Lower is better.}
\label{tab:phylogfn_phylaflow_mcmc_treekl}
\begin{tabular}{lcc}
\toprule
Dataset & PhyloGFN-MCMC & PhylaFlow-MCMC \\
\midrule
DS3 & \textbf{0.047422} & 0.127418 \\
DS5 & 2.045933 & \textbf{1.661833} \\
DS6 & 9.218921 & \textbf{4.564091} \\
DS7 & 3.854252 & \textbf{1.070613} \\
\bottomrule
\end{tabular}
\end{table}

\subsection{\name Conditioning Results}
\label{Appendix:condit:phyla}

We utilized Phyla embeddings because previous work~\cite{ektefaie2026evolutionary} has shown that these embeddings capture evolutionary relationships, specifically phylogenetic tree structure, better than existing protein language models. Since Phyla was only trained on proteins, we trained a variant of Phyla on both proteins and nucleic acids, utilizing the orthomam~\cite{Ranwez2007-gz} dataset and following the training procedure described in the paper.

\paragraph{Per-taxon sequence conditioning \(\Phi(S)\).} The model
    implements two paths that inject per-taxon sequence embeddings into the
    encoder input: a leaf-token addition \(x_i^{\mathrm{leaf}}\!\leftarrow\!x_i^{\mathrm{leaf}}+W_{\mathrm{leaf}} z_i\)
    and a split-bipartition MLP added to each internal-edge token. For the
    DS1--DS8 results reported in this paper, both paths are disabled
    and the model is conditioned only on the starting tree through the frozen case embedding.

\begin{table}[t]
\centering
\small
\setlength{\tabcolsep}{6pt}
\renewcommand{\arraystretch}{1.06}
\caption{\textbf{Joint DS1--DS8 conditioning results.}
A single checkpoint is trained jointly across DS1--DS8 using dataset-specific
Phyla embedding banks. We report the number of sampled starts, the split-KL and
Tree-KL of the native conditioned samples, and the final Tree-KL after
$10^5$ MrBayes generations initialized from those samples.}
\label{Appendix:condit:kl}
\begin{tabular}{lrrrr}
\toprule
Dataset & Starts & Sample split-KL & Sample Tree-KL & MB $10^5$ final Tree-KL \\
\midrule
DS1 & 234  & 0.170466 & 12.364656 & 0.418221 \\
DS2 & 42   & 0.083351 & 6.418691  & 0.066534 \\
DS3 & 243  & 0.077111 & 9.737936  & 2.550395 \\
DS4 & 573  & 0.291568 & 16.980017 & 0.283246 \\
DS5 & 525  & 0.119899 & 14.936400 & 3.697866 \\
DS6 & 219  & 0.188916 & 16.852033 & 5.140047 \\
DS7 & 1344 & 0.105480 & 14.259826 & 2.543646 \\
DS8 & 1122 & 0.271230 & 16.406969 & 0.856555 \\
\bottomrule
\end{tabular}
\end{table}

\begin{table}[t]
\centering
\small
\setlength{\tabcolsep}{5pt}
\renewcommand{\arraystretch}{1.06}
\caption{\textbf{Cross-conditioning stress test for the jointly trained checkpoint.}
For each target dataset, the checkpoint, starts, posterior reference, and scoring
procedure are fixed; only the Phyla embedding bank is changed. Native conditioning
is the diagonal condition. Off-diagonal swaps are included only when leaf-position
compatibility permits. Lower split-KL is better.}
\label{Appendix:condit:crosstable}
\begin{tabular}{lrrrr}
\toprule
Target & Native split-KL & Best valid swap & Best swap split-KL & Native gap \\
\midrule
DS1 & 0.473 & DS6 & 3.475 & 3.003 \\
DS2 & 0.126 & DS6 & 3.101 & 2.975 \\
DS3 & 0.087 & DS7 & 3.507 & 3.419 \\
DS4 & 0.392 & DS6 & 3.539 & 3.148 \\
DS5 & 0.185 & DS7 & 4.036 & 3.851 \\
DS6 & 0.176 & DS7 & 3.623 & 3.447 \\
DS7 & 0.215 & DS8 & 3.889 & 3.674 \\
DS8 & 0.278 & none valid & -- & -- \\
\midrule
Mean over evaluable targets & 0.236 & -- & 3.596 & 3.359 \\
\bottomrule
\end{tabular}
\end{table}

\subsection{Ablations}
\label{Appendix:ablations}

To assess the contribution of each component, we performed ablations on the DS1-trained model for computational efficiency. Removing any component degraded performance, indicating that each part contributes to accurate tree prediction. The full model achieved the lowest Tree-KL of 2.728 (Table~\ref{tab:ds1_ablation_tree_kl}). Among the ablations, removing the velocity head caused a moderate increase in Tree-KL, while removing start-tree conditioning or the first-hit head led to substantially larger degradation. The largest drop occurred when removing the AR-head, increasing Tree-KL by 13.813 over the full model, suggesting that autoregressive conditioning is the most critical component for modeling tree structure accurately.

\begin{table}[t]
  \centering
  \caption{DS1 ablation results. Lower tree-KL is better.}
  \label{tab:ds1_ablation_tree_kl}
  \begin{tabular}{l r r}
  \toprule
  Variant & Tree-KL & $\Delta$ vs. full \\
  \midrule
  Full model & 2.728 & 0.000 \\
  No velocity head & 6.282 & +3.554 \\
  No start tree conditioning & 9.634 & +6.906 \\
  No first-hit head & 12.097 & +9.369 \\
  No AR-head & 16.541 & +13.813 \\
  \bottomrule
\end{tabular}
\end{table}


\end{document}